\RequirePackage{fix-cm}
\documentclass[smallextended]{svjour3}       
\smartqed  
\usepackage{graphicx}
\usepackage{amsmath}

\usepackage{lineno}

\usepackage{natbib}
\setcitestyle{aysep={}}

\usepackage{url}

 \journalname{Meteorology and Atmospheric Physics}
\begin{document}

\title{Modelling of Vortex-Induced Aviation Turbulence}


\author{Nils T. Basse}


\institute{   Elsas v\"ag 23, 423 38 Torslanda, Sweden \\
              Tel.: +46-705200362\\
              \email{nils.basse@npb.dk}}

\date{Received: date / Accepted: date}

\maketitle

\begin{abstract}
Aviation turbulence is modelled as an interaction between an aircraft and a vortex tube. The vortex tube can have an arbitrary orientation/offset with respect to the aircraft. We compare modelling the aircraft (i) as a point and (ii) having a finite area (wing and fuselage). We consider both vertical and horizontal acceleration experienced by the aircraft. The baseline vortex tube has an area which is of the order of the aircraft area.
\end{abstract}

\section{Introduction}

Aviation turbulence \citep{sharman_a} has an impact on flight comfort, safety and cost. A better understanding - and modelling - of the physical phenomena is needed to reduce the undesired effects of aviation turbulence. In this paper, we will assume that (a part of) aviation turbulence can be described as an interaction between an aircraft (AC) and a vortex tube (VT) \citep{lamb_a,saffman_a}. Our starting point is \citep{lunnon_a}.

Illustrations of how VTs can form are shown in e.g. Fig. 1 in \citep{wingrove_a}. The underlying instabilities are described in \citep{lin_a}. One candidate instability, the Kelvin-Helmholtz instability, is described in more detail in \citep{batchelor_a} as "The instability of a sheet vortex".

Horizontal vortex tubes (HVTs) have been identified to occur during aviation turbulence \citep{clark_a}. A physical mechanism for the formation of HVTs has been provided in \citep{roach_a,kaplan_a}.

The main purpose of our paper is to extend the modelling efforts in \citep{parks_a,mehta_a}: In those papers, HVTs were modelled and the aircraft was assumed to be a point. Our extension includes VTs of arbitrary orientation/offset and a finite area of the aircraft (wing/fuselage). Further, we consider both vertical (normal) and horizontal (transverse) acceleration of the aircraft.

Traditionally the normal component of acceleration is regarded as the most important from the perspective of the safety of the aircraft - large normal accelerations can result in aircraft damage. From the perspective of passenger comfort, horizontal accelerations can cause injuries; this is reflected in the formulation of the Dose of Discomfort measure \citep{jacobson_a} in which all three components of acceleration are given equal weight.

\section{Modelling}

\subsection{Vortex Tube Properties}

\begin{figure}
\includegraphics[width=10cm]{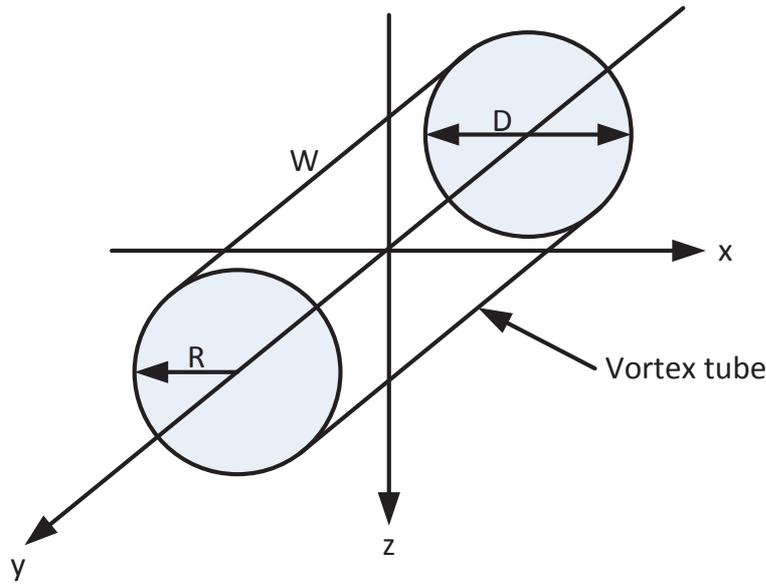}
\caption{Baseline vortex tube geometry and dimensions.}
\label{fig:baseline_vortex}
\end{figure}

The vortex is modelled as a circular 2D vortex extending into the third dimension. The baseline vortex is shown in Fig. \ref{fig:baseline_vortex}. The vortex radius is $R$, the vortex diameter is $2R=D$ and the vortex width is $W$. We note that a finite vortex width is not supported by theory, which states that vortex tubes either extend to infinity or end on solid boundaries \citep{saffman_a}. The vortex axis is along the $y$-direction.

\begin{figure}
\includegraphics[width=10cm]{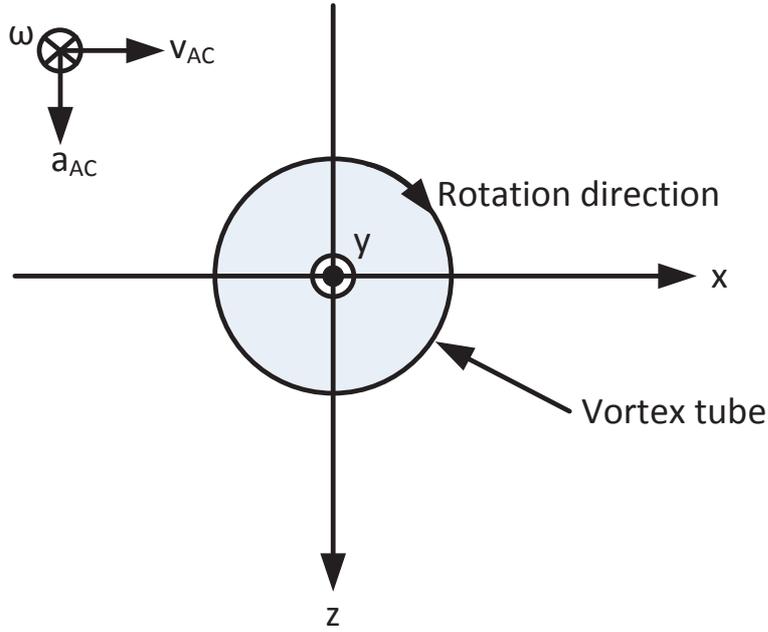}
\caption{Vortex tube velocity direction and resulting aircraft acceleration.}
\label{fig:aircraft_acc}
\end{figure}

The aircraft travels in the $x$-direction. The clockwise direction of the vortex velocity in the $xz$-plane leads to vorticity in the negative $y$-direction, see Fig. \ref{fig:aircraft_acc}.

With respect to the aircraft, the $x$-direction is longitudinal, the $y$-direction is transverse and the $z$-direction is vertical. Since we will see that the acceleration is zero in the longitudinal direction, we will use the term horizontal for the transverse direction ($y$).

We assume that the vortex rotates like a solid body with a constant angular velocity $\Omega$:

\begin{equation}
\boldsymbol\Omega=\left( 0,-\Omega,0\right)
\end{equation}

The vortex radius is $\boldsymbol r=\left( x,0,z\right)$. The resulting vortex velocity is:

\begin{equation}
\boldsymbol u=\boldsymbol\Omega \times \boldsymbol r=\left(-\Omega z,0,\Omega x\right)
\end{equation}

The vorticity (vortex strength) is:

\begin{equation}
\boldsymbol\omega=\nabla \times \boldsymbol u=\left( 0,-2\Omega,0\right)=2 \boldsymbol\Omega
\end{equation}

For the general case we define the total vorticity as:

\begin{equation}
\omega_{\rm tot} \equiv \sqrt{\omega_x^2 + \omega_y^2 + \omega_x^2}
\end{equation}

For the specific case, we have $\boldsymbol\omega=\left(0,\omega_y,0 \right)=\left( 0,-2\Omega,0\right)$ and $\omega_{\rm tot}=|\omega_y|=2\Omega$.

The vortex speed is:

\begin{equation}
u=r\Omega,
\end{equation}

\noindent where $r$ is the radius inside the vortex (from 0 to $R$).

The vortex circulation for radius $r$ is \citep{gerz_a}:

\begin{equation}
\Gamma(r)=\omega_{\rm tot} r^2 \pi =  2 \Omega r^2 \pi,
\end{equation}

\noindent so the total circulation of the vortex is:

\begin{equation}
\Gamma_{\rm tot} = \Gamma(R) = 2 \Omega R^2 \pi
\end{equation}

The vortex lifetime can be approximated by the eddy-turnover time \citep{pope_a}: Using $D$ as the length scale and $u(r=R)$ as the velocity, we get:

\begin{equation}
\tau_{\rm vortex}=D/u(r=R)=2/\Omega
\end{equation}

\subsection{Aircraft Acceleration}

The main quantity we study is the aircraft acceleration due to a vortex tube as defined by Lunnon \citep{lunnon_a}:

\begin{equation}
{\bf a}_{\rm AC} = \frac{1}{2} \boldsymbol\omega \times {\bf v_{\rm AC}},
\label{eq:lunnon_acc}
\end{equation}

\noindent where ${\bf v_{\rm AC}}$ is the aircraft velocity, see Fig. \ref{fig:aircraft_acc}.

Eq. (\ref{eq:lunnon_acc}) can be derived from the fluid \citep{lamb_a,batchelor_a} (i.e. hydrodynamic \citep{saffman_a}) impulse of the aircraft:

\begin{equation}
{\bf I_{\rm AC}} = - \frac{1}{2} \rho_{\rm AC} \int {\bf r_{\rm AC}} \times \boldsymbol\omega {\rm d} V_{\rm AC},
\label{eq:impulse}
\end{equation}

\noindent where ${\bf r_{\rm AC}}$ is the aircraft position, $\rho_{\rm AC}=\frac{m_{\rm AC}}{V_{\rm AC}}$ is the aircraft density, $m_{\rm AC}$ is the aircraft mass and $V_{\rm AC}$ is the aircraft volume. Note that $m_{\rm AC}$ and $V_{\rm AC}$ are defined for the part of the aircraft which is enclosed by the vortex.

Here, we have modified the original expression in two ways:
\begin{itemize}
  \item We have introduced a negative sign to account for the fact that we study the impulse of the aircraft instead of the impulse of the vortex
  \item We have replaced vortex by aircraft quantities where relevant
\end{itemize}

We assume that the vorticity of the vortex is independent of time. Thus, we can take the time derivative of the impulse to arrive at the force on the aircraft:

\begin{eqnarray}
  {\bf F_{\rm AC}} = \frac{{\rm d}{\bf I_{\rm AC}}}{{\rm d}t} = m_{\rm AC} {\bf a}_{\rm AC} &=& - \frac{1}{2} \frac{m_{\rm AC}}{V_{\rm AC}} \int {\bf v_{\rm AC}} \times \boldsymbol\omega {\rm d} V_{\rm AC} \\
  {\bf a}_{\rm AC} &=& \frac{1}{V_{\rm AC}} \int \frac{1}{2} \boldsymbol\omega \times {\bf v_{\rm AC}} {\rm d} V_{\rm AC}
\end{eqnarray}

Since the aircraft velocity is constant, we can remove the integral; this concludes our derivation of Eq. (\ref{eq:lunnon_acc}).

We assume that the vortex is stationary and that the aircraft is moving. If the vortex is moving, the aircraft velocity is interpreted as the velocity difference between the aircraft and the vortex.

We assume that the acceleration will not move the aircraft, i.e. the aircraft trajectory is fixed.

Using Eq. (\ref{eq:lunnon_acc}) and the fact that the aircraft is travelling in the $x$-direction we find:

\begin{equation}
{\bf a}_{\rm AC} = \frac{1}{2} \left(0,-2\Omega,0 \right) \times \left(v_{\rm AC},0,0\right) = \left(0,0,\Omega v_{\rm AC} \right)
\end{equation}

Thus, the aircraft experiences an acceleration in the positive $z$-direction, see Fig. \ref{fig:aircraft_acc}.

\subsection{Example}
\label{subsec:example}

We use a typical cruising speed $v_{\rm AC} = 800$ km/h = 800/3.6 m/s = 222.2 m/s.

From \citep{sharman_b}, we set the aircraft acceleration of a significant event to $a_{\rm AC} = 0.5$ g = 4.9 m/s$^2$.

Combining the above, the angular velocity of the vortex $\Omega = a_{\rm AC}/v_{\rm AC} = 2.2 \times 10^{-2}$ s$^{-1}$. The resulting vorticity $\omega_y = -2\Omega = -4.4 \times 10^{-2}$ s$^{-1}$.

We choose a vortex radius $R = 10.7$ m based on the discussion in Section \ref{subsubsec:p_vs_a}.

This leads to a maximum speed on the vortex surface $u(r=R) = R \Omega$ = 0.24 m/s.

The total circulation of the vortex is $\Gamma_{\rm tot}$ = 15.9 m$^2$/s.

The eddy-turnover time $D/u(r=R)=2/\Omega= 91$ s. This can be compared to the duration when the aircraft is inside the vortex $D/v_{\rm AC} = 0.1 $ s. For this case, the vortex persists much longer than the aircraft takes to pass through it, so we do not have to consider vortex decay. However, vortex lifetime will become increasingly important for larger $\Omega$ and/or slower aircraft.

\subsection{Aircraft Geometry}

\subsubsection{Point}

The simplest aircraft model is to consider it as a point. The implicit assumption is that the vortex dimensions are much larger than the aircraft dimensions. We will discuss the errors using this approach in more detail below.

\subsubsection{Area}

When we model the aircraft with a finite area, we use dimensions approximating those of an Airbus A330-200 \citep{airbus_a}, i.e. a medium-sized aircraft.

\begin{figure}
\includegraphics[width=5cm]{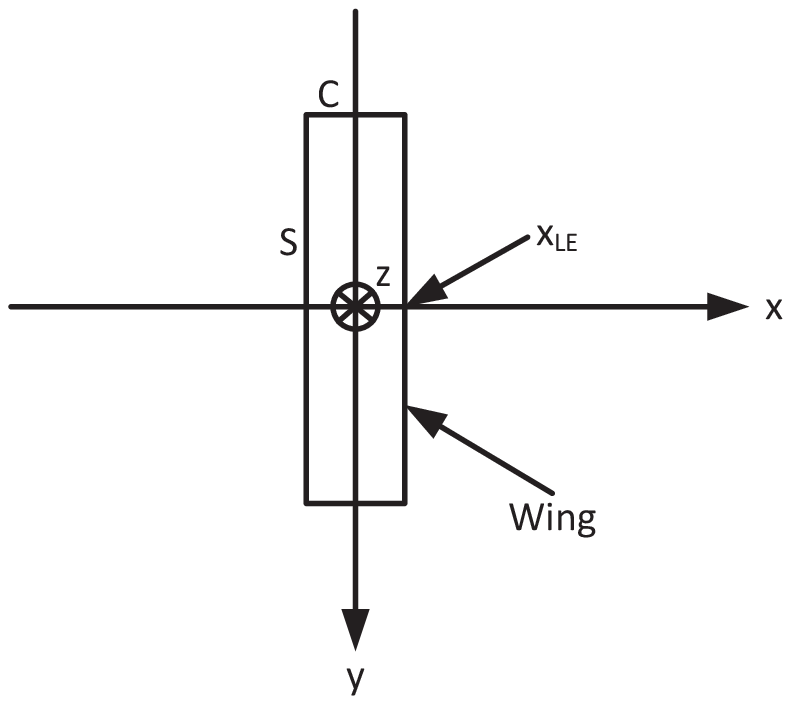}
\includegraphics[width=5cm]{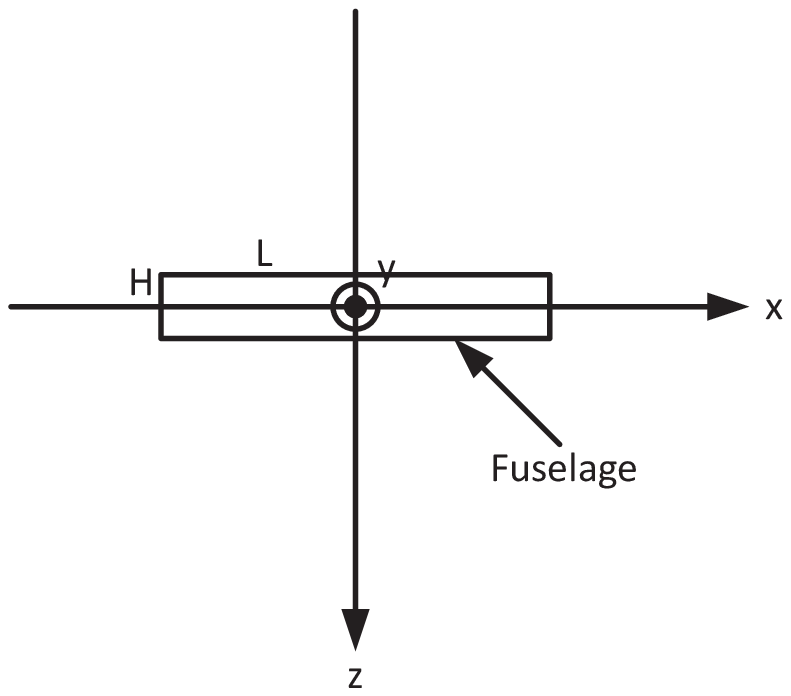}
\caption{Aircraft geometry: Wing (left) and fuselage (right).}
\label{fig:aircraft_wing_fuselage}
\end{figure}

The wing span $S$ is fixed at 60 m and the wing chord $C$ is set to 6 m, see Fig. \ref{fig:aircraft_wing_fuselage}. The position of the leading edge (LE) of the wing is shown as $x_{\rm LE}$. The fuselage length $L$ is 60 m and the fuselage height $H$ is 6 m.

The wing is assumed to be fixed exactly at the center of the fuselage. Both the wing and fuselage are assumed to be rectangles. The horizontal stabilizer and tail fin are not considered.

It would be more realistic to represent the fuselage as a cylinder rather than a rectangle - the cross section of the fuselage of an Airbus A330-200 is essentially circular. This could be achieved by adding a second horizontal rectangle, coaxial with the vertical rectangle which represents the fuselage, in the computations. However, the approach used is justified by being the simplest aircraft representation for which both normal and transverse accelerations can be derived.

\subsubsection{Point Versus Area}
\label{subsubsec:p_vs_a}

To discuss the comparison between the vortex and aircraft areas, we define the vortex/aircraft area ratio $r_A$. For the fuselage, the vortex $xz$-plane area is $\pi R^2$ and the aircraft fuselage area is $LH$. For a fixed $r_A$, the vortex radius $R_A$ is given:

\begin{eqnarray}
r_A &=& \frac{\pi R_A^2}{LH} \\
R_A &=& \sqrt{\frac{r_A L H}{\pi}}
\label{eq:R_A}
\end{eqnarray}

As for the fuselage, we also require that $r_A$ is equal to the ratio of the vortex $xy$-plane area $DW=2RW$ and the aircraft wing area $SC$:

\begin{eqnarray}
  r_A &=& \frac{2 R_A W_A}{S C} \\
  W_A &=& \frac{r_A S C}{2 R_A}
  \label{eq:W_A}
\end{eqnarray}

Combining Eqs. (\ref{eq:R_A}) and (\ref{eq:W_A}), we write the vortex width $W_A$ as:

\begin{equation}
  W_A = \frac{S C}{2} \sqrt{\frac{r_A \pi}{L H}}
\end{equation}

\begin{figure}
\vspace{0.5cm}
\includegraphics[width=10cm]{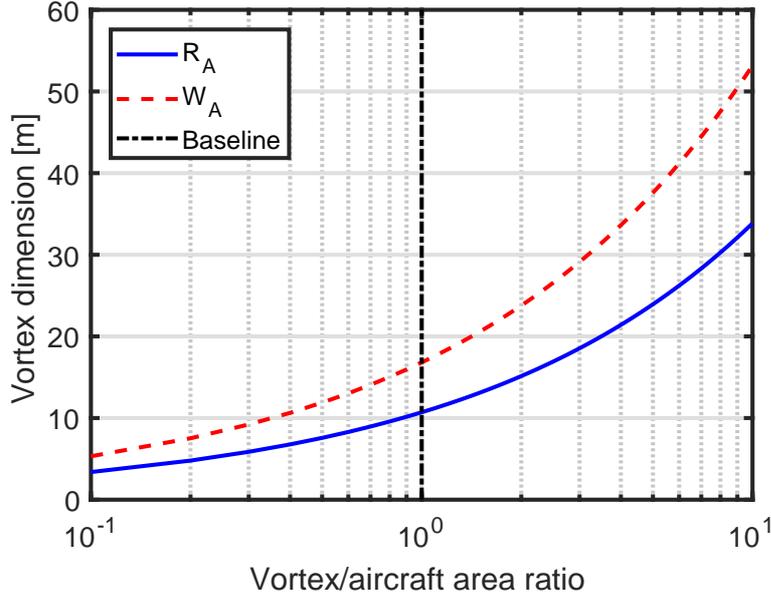}
\caption{Vortex dimensions as a function of vortex/aircraft area ratio.}
\label{fig:vortex_ratio}
\end{figure}

We can now use $r_A$ to determine the area ratio; this is illustrated in Fig. \ref{fig:vortex_ratio} for the aircraft dimensions provided above, where $R_A$ and $W_A$ are shown as a function of $r_A$. We will consider three cases below:

\begin{itemize}
  \item $r_A=0.1$ ($R=3.4$ m, $W=5.3$ m): Vortex area is much smaller than aircraft area
  \item $r_A=1$ ($R=10.7$ m, $W=16.8$ m): Vortex area is the same as the aircraft area (our baseline case)
  \item $r_A=10$ ($R=33.9$ m, $W=53.2$ m): Vortex area is much larger than aircraft area
\end{itemize}

For the first two cases, the aircraft should be modelled with finite areas. For the last case, modelling the aircraft as a point is sufficient for most purposes.

\section{Results: Aircraft as a Point}
\label{sec:ac_as_pt}

Here, we discuss analytical modelling of the aircraft as a point.

\subsection{Vortex Offset}
\label{subsec:vt_offset}

The vortex can be offset in the horizontal direction $y_0$. When $-W/2 < y_0 < W/2$, the aircraft will be inside the vortex, otherwise it will be outside.

For a vertical offset $z_0$, the aircraft will travel a distance $2 \sqrt{R^2 - z_0^2}$ inside the vortex. For example, if we want to reduce the distance from $2R$ to $R$, we set $z_0 = R \sqrt{\frac{3}{4}}$.

\subsection{Vortex Angle}

\subsubsection{Left-Right Tilt}

\begin{figure}
\includegraphics[width=3.5cm]{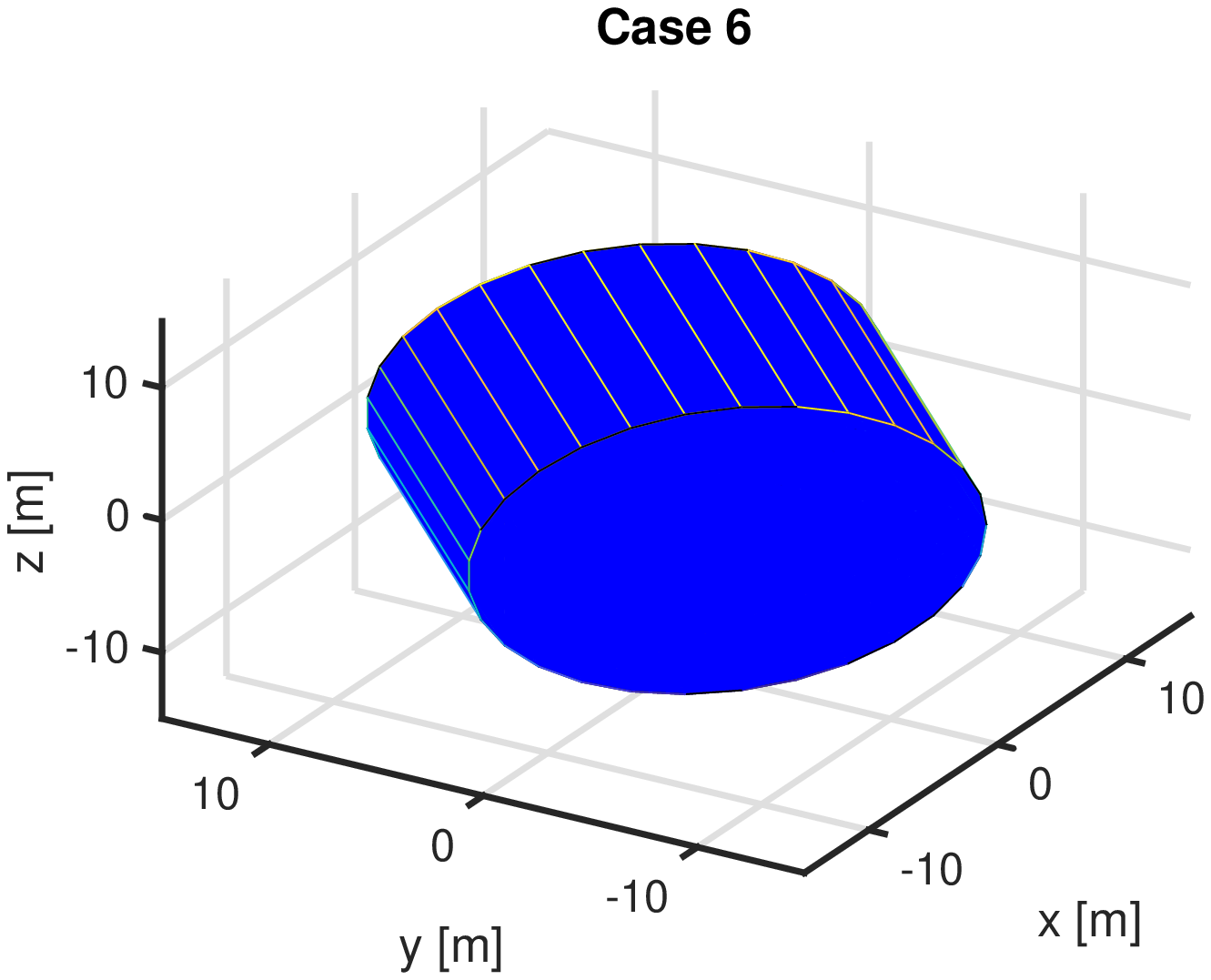}
\hspace{0.5cm}
\includegraphics[width=3.5cm]{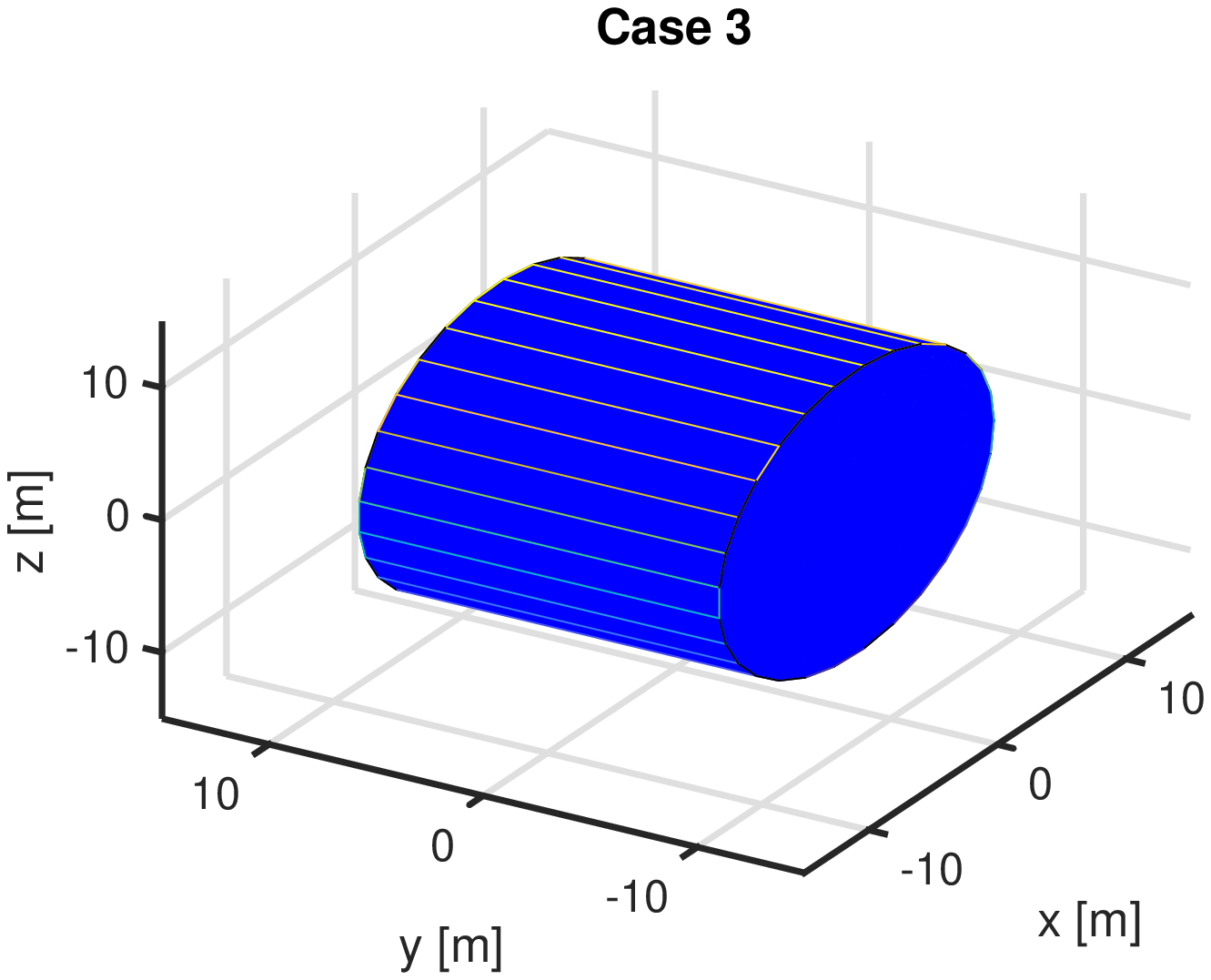}
\hspace{0.5cm}
\includegraphics[width=3.5cm]{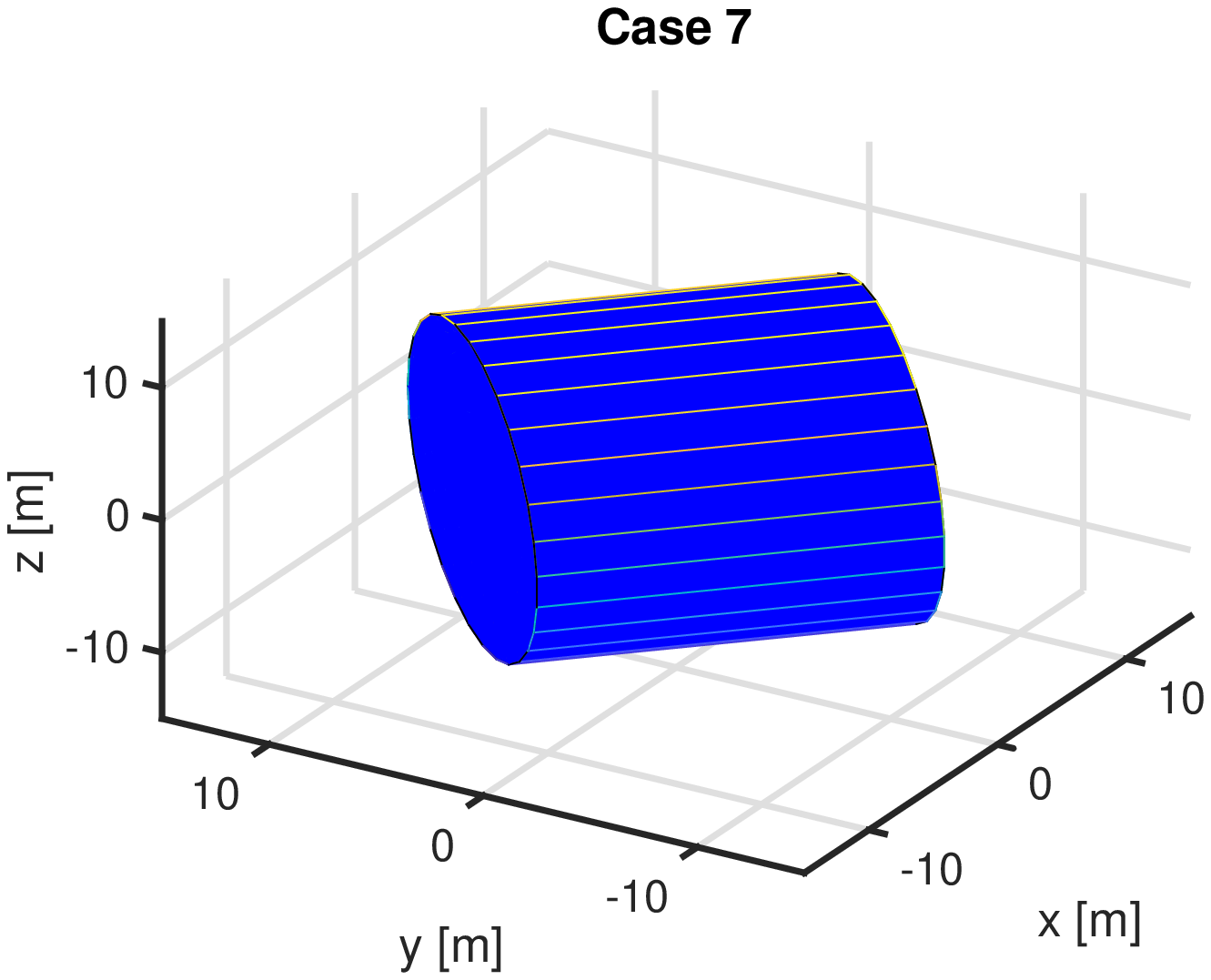}
\caption{Vortex tube orientation for left-right tilt: $\phi=\pi/4$ (left), $\phi=\pi/2$ (center), $\phi=3\pi/4$ (right).}
\label{fig:phi_cylinder}
\end{figure}

A left-right tilt of the vortex is rotation in the $xy$-plane. This is defined using the azimuthal angle $\phi$, where $\pi/2$ is the baseline case. See Fig. \ref{fig:phi_cylinder} for an illustration of the corresponding vortex tube orientation.

The aircraft will travel a distance $2R/\sin(\phi)$ inside the vortex.

The vorticity is given by:

\begin{eqnarray}
  \boldsymbol\omega &=& \left(\omega_x,\omega_y,0 \right) \\
  \omega_x &=& -2 \Omega \cos \phi\\
  \omega_y &=& -2 \Omega\sin \phi
\end{eqnarray}

The resulting aircraft acceleration is in the vertical direction:

\begin{equation}
{\bf a}_{\rm AC} = \frac{1}{2} \left( \omega_x,\omega_y,0 \right) \times \left(v_{\rm AC},0,0\right) = \frac{1}{2} \left( 0,0,-\omega_y v_{\rm AC} \right)
\end{equation}

\subsubsection{Up-Down Tilt}

\begin{figure}
\includegraphics[width=3.5cm]{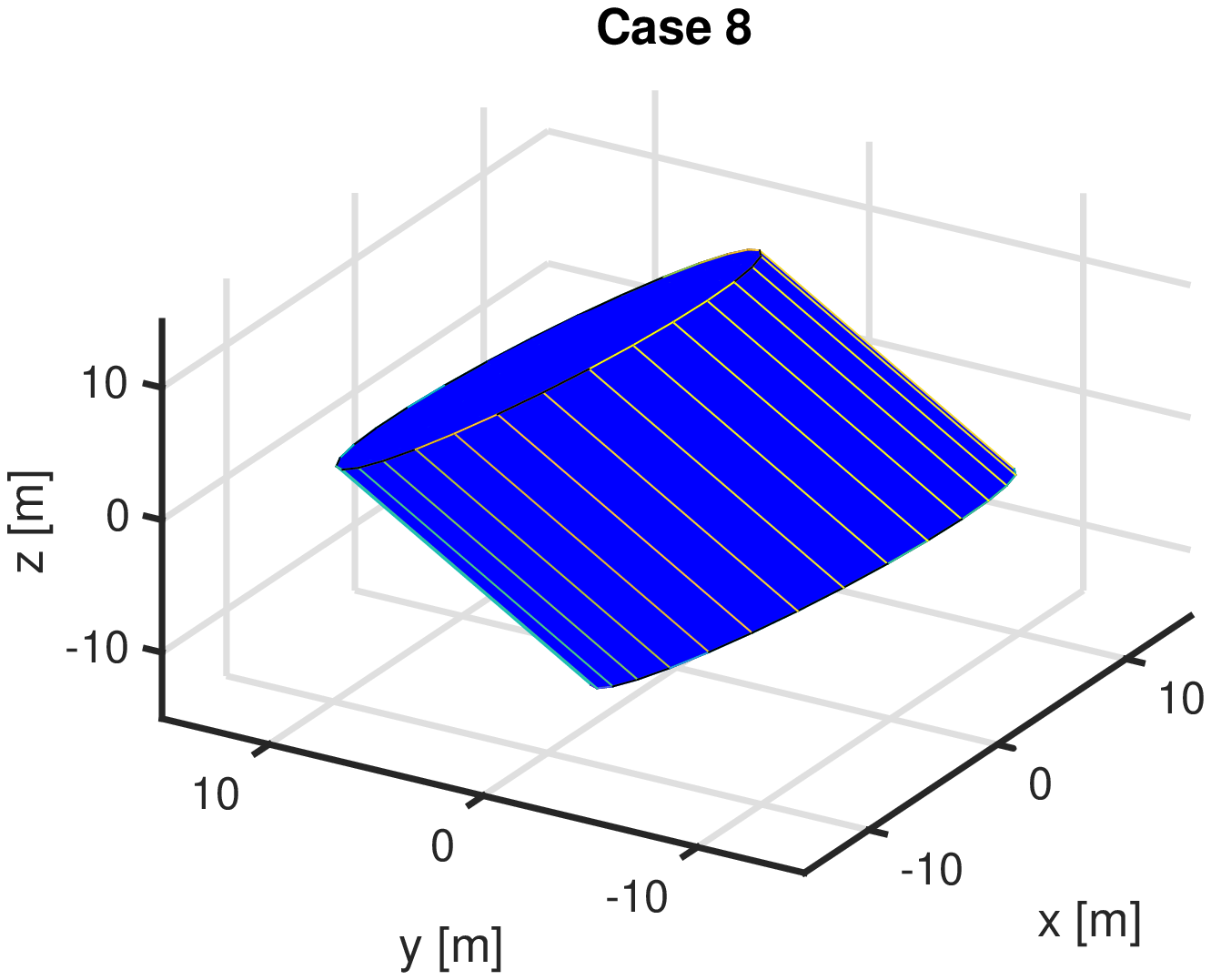}
\hspace{0.5cm}
\includegraphics[width=3.5cm]{case_3_cylinder.eps}
\hspace{0.5cm}
\includegraphics[width=3.5cm]{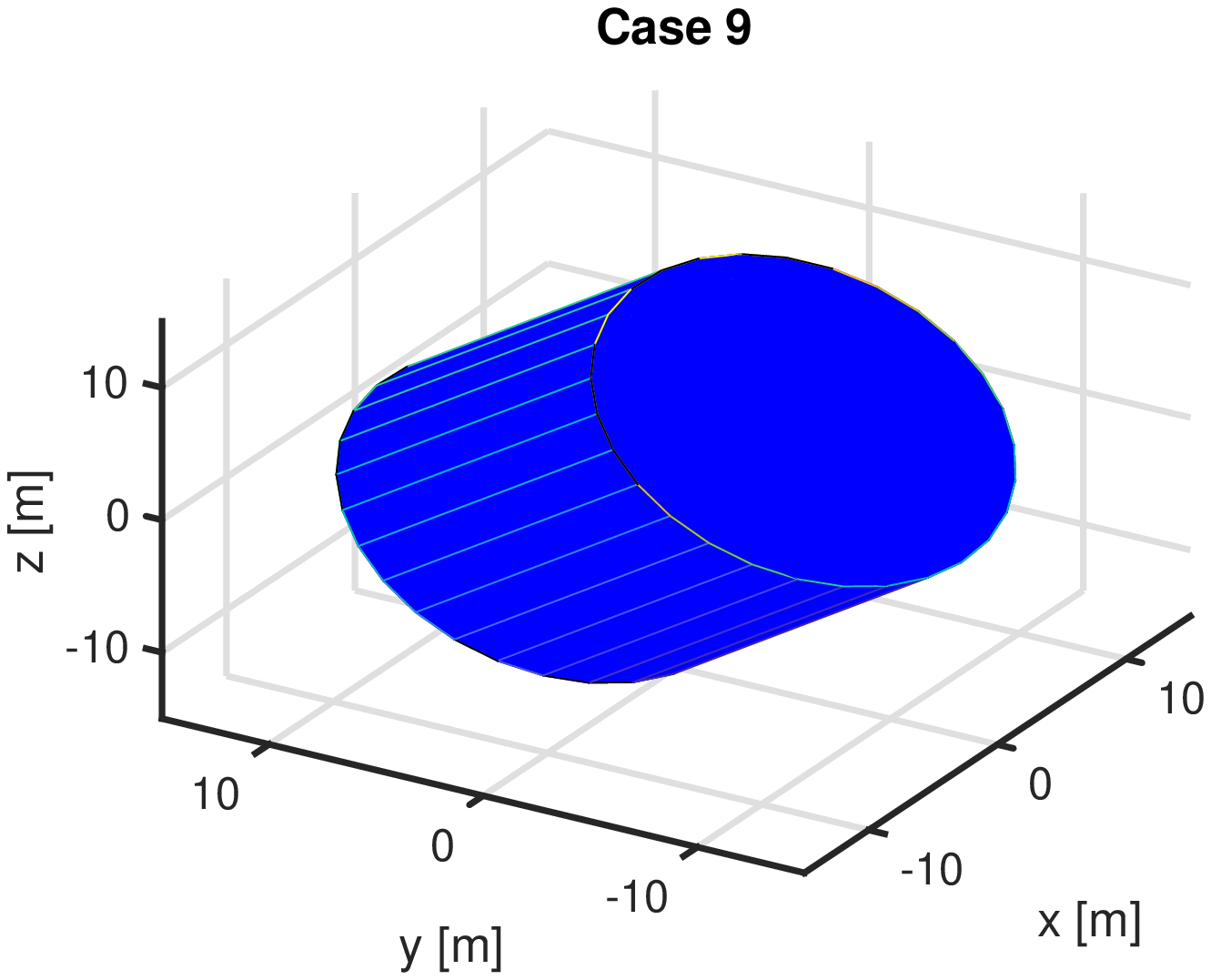}
\caption{Vortex tube orientation for up-down tilt: $\theta=\pi/4$ (left), $\theta=\pi/2$ (center), $\theta=3\pi/4$ (right).}
\label{fig:theta_cylinder}
\end{figure}

An up-down tilt of the vortex is rotation in the $xz$-plane. This is defined using the polar angle $\theta$, where $\pi/2$ is the baseline case. See Fig. \ref{fig:theta_cylinder} for an illustration of the corresponding vortex tube orientation.

The aircraft will travel a distance $2R$ inside the vortex.

The vorticity is given by:

\begin{eqnarray}
  \boldsymbol\omega &=& \left(0,\omega_y,\omega_z \right) \\
  \omega_y  &=& -2 \Omega \sin \theta \\
  \omega_z &=& -2 \Omega \cos \theta
\end{eqnarray}

The resulting aircraft acceleration is in both the horizontal and the vertical direction:

\begin{equation}
{\bf a}_{\rm AC} = \frac{1}{2} \left( 0,\omega_y,\omega_z \right) \times \left(v_{\rm AC},0,0\right) = \frac{1}{2} \left( 0,\omega_zv_{\rm AC},-\omega_y v_{\rm AC} \right)
\end{equation}

\subsubsection{Combined Tilt}

For a combined tilt, both $\phi$ and $\theta$ are allowed to vary between $0$ and $\pi$.

The vorticity is given by:

\begin{eqnarray}
  \omega_x &=& -2 \Omega \sin \theta \cos \phi \\
  \omega_y &=& -2 \Omega \sin \theta \sin \phi\\
  \omega_z &=& -2 \Omega \cos \theta
\end{eqnarray}

The resulting aircraft acceleration is in both the horizontal and the vertical direction:

\begin{equation}
{\bf a}_{\rm AC} = \frac{1}{2} \left( \omega_x,\omega_y,\omega_z \right) \times \left(v_{\rm AC},0,0\right) = \frac{1}{2} \left( 0,\omega_zv_{\rm AC},-\omega_y v_{\rm AC} \right)
\end{equation}

\begin{figure}
\includegraphics[width=5cm]{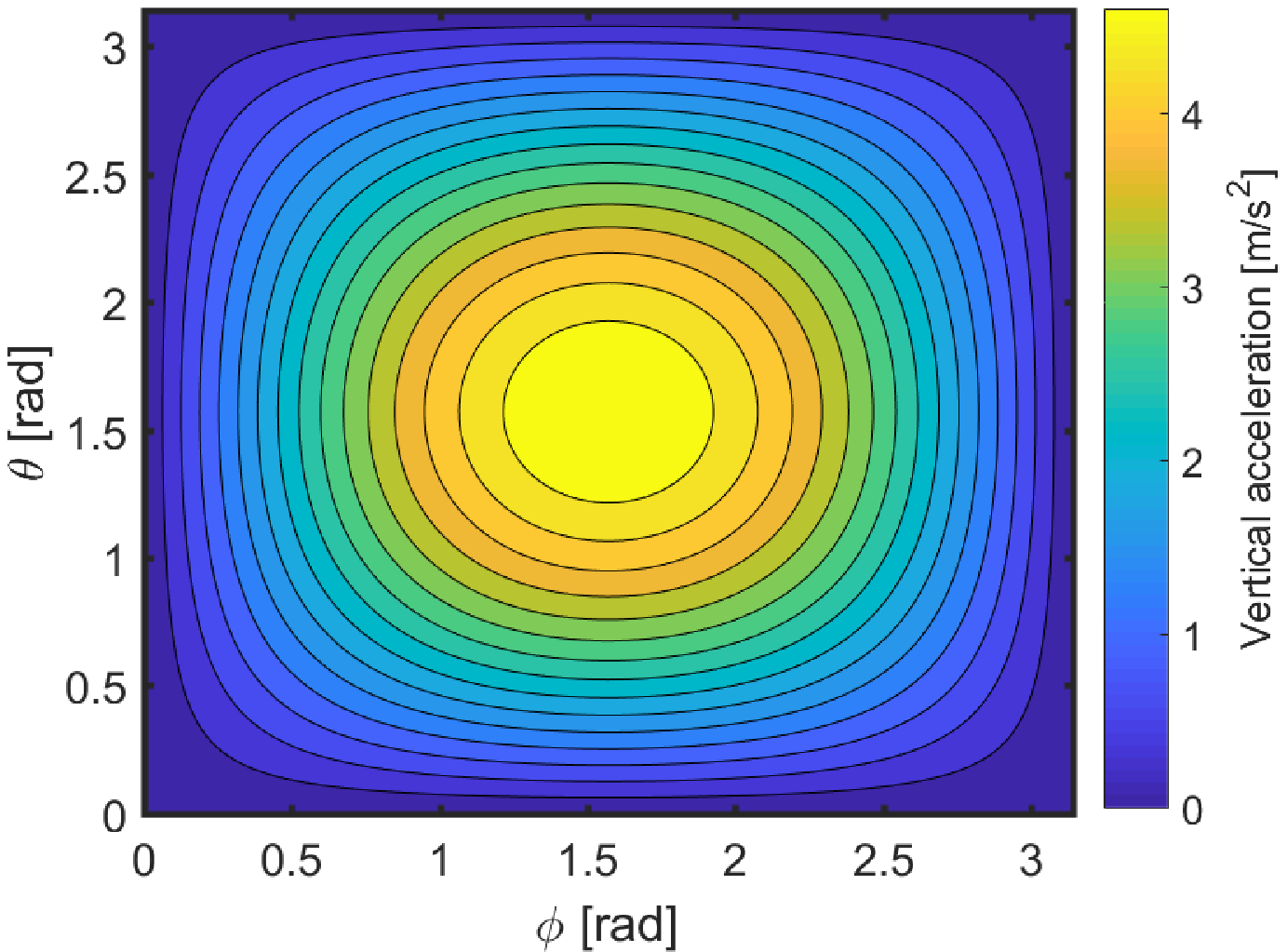}
\includegraphics[width=5cm]{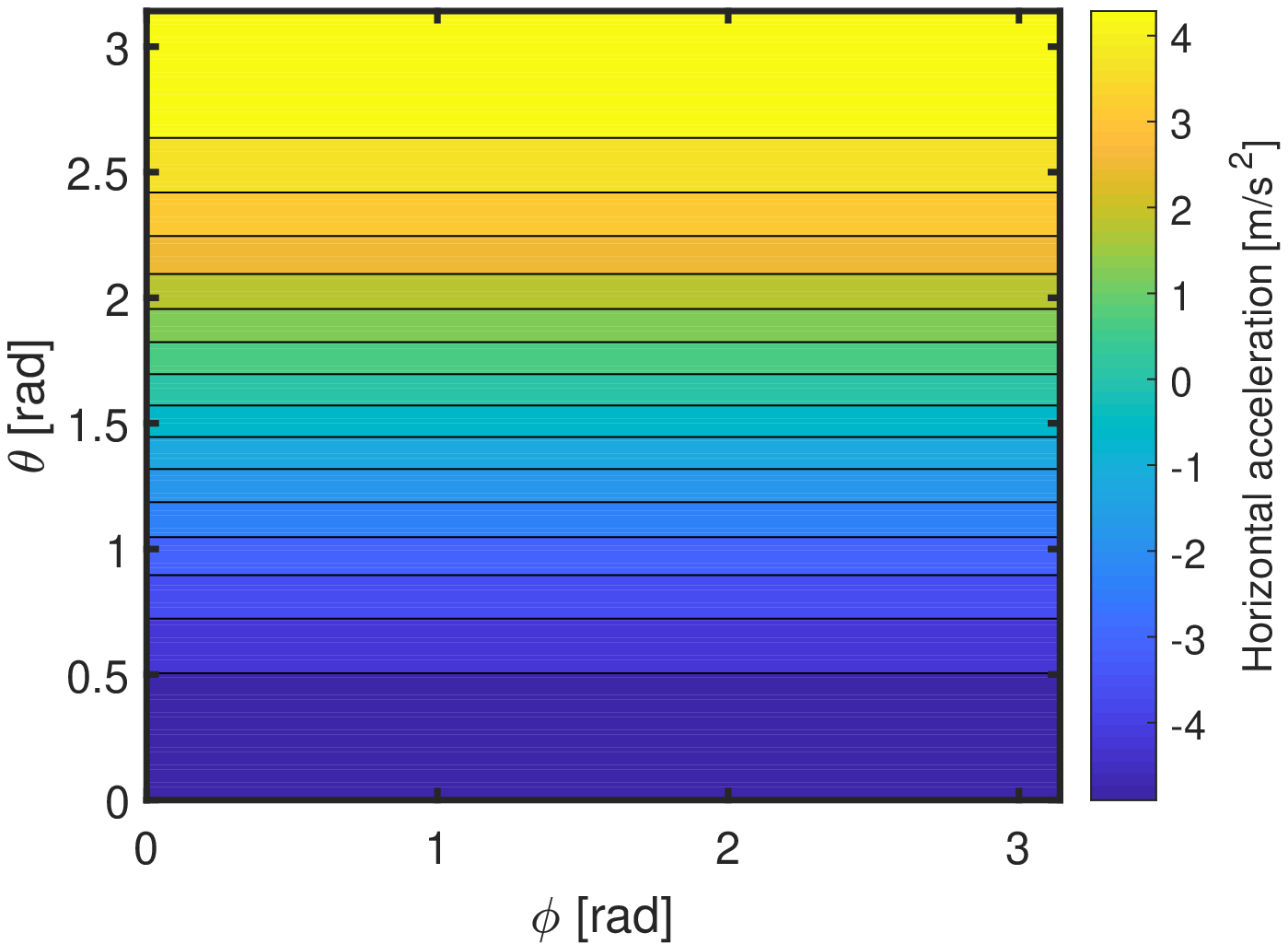}
\caption{Left: Vertical acceleration, right: Horizontal acceleration.}
\label{fig:cont_vert_hor_acc}
\end{figure}

For values from our example above, the acceleration in the vertical and horizontal direction is shown as contour plots in Fig. \ref{fig:cont_vert_hor_acc}.

The sign of the vertical acceleration is always positive and is symmetric around the baseline orientation $(\phi,\theta)=(\pi/2,\pi/2)$ (which is the maximum acceleration).

The horizontal acceleration has a positive maximum for $\theta=\pi$ and a negative minimum for $\theta=0$. The horizontal acceleration is independent of $\phi$.

\section{Results: Aircraft as an Area}

\subsection{Numerical Procedure}

The numerical procedure when considering the aircraft to be an area consists of two steps. The first step is the generation of the vortex tube:

\begin{enumerate}
\item Create VT (split in 30 segments)
\item Apply offsets $y_0$ and $z_0$
\item Calculate convex hull of VT
\item Check that the volume of the convex hull is the same as the VT volume
\item Define grids for the $xy$- and $xz$-planes (resolution 0.5 m)
\item Calculate intersection between the VT and the two planes
\item Calculate the convex hull of the VT in the two planes
\end{enumerate}

Once the VT is defined, the position of the aircraft is a function of time. We use 71 timesteps with a step length of 0.01 s, i.e. a total duration of 0.7 s. Using the aircraft speed from our example in Section \ref{subsec:example}, this is equivalent to a distance of 156 m.

The second step proceeds as follows:

\begin{enumerate}
\item Define $x_{\rm LE}$ for the given timestep
\item Define both the wing and the fuselage as four point polygons
\item Define the aircraft center point
\item Calculate the intersection between the VT in the $xy$-plane ($xz$-plane) and the wing (fuselage), respectively
\item Check if center point is inside the VT
\item Calculate the ratio between the intersection area and the total wing (fuselage) area, respectively
\item Multiply the maximum acceleration (0.5 g) by the area ratios
\end{enumerate}

\begin{figure}
\includegraphics[width=5cm]{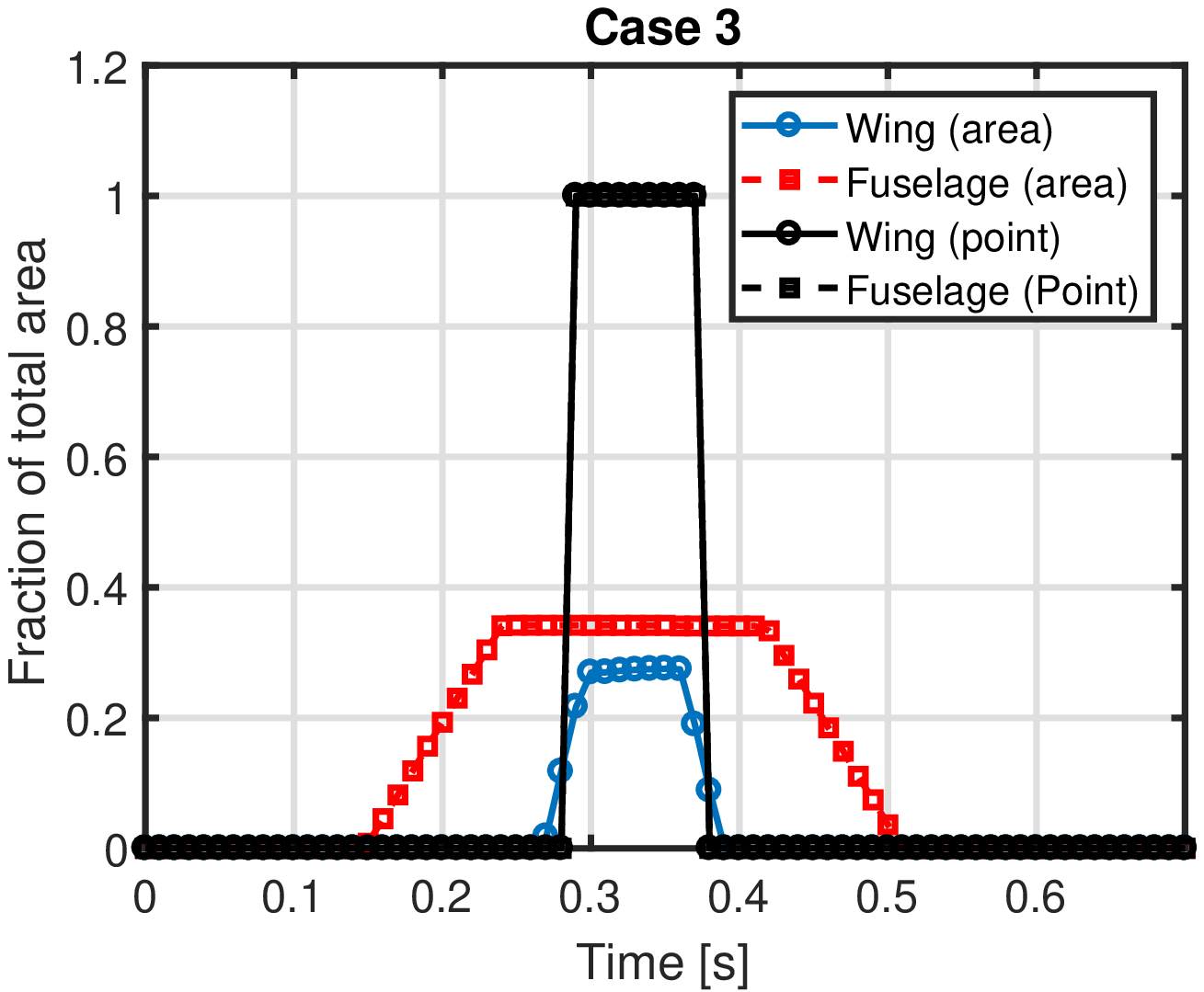}
\hspace{0.5cm}
\includegraphics[width=5cm]{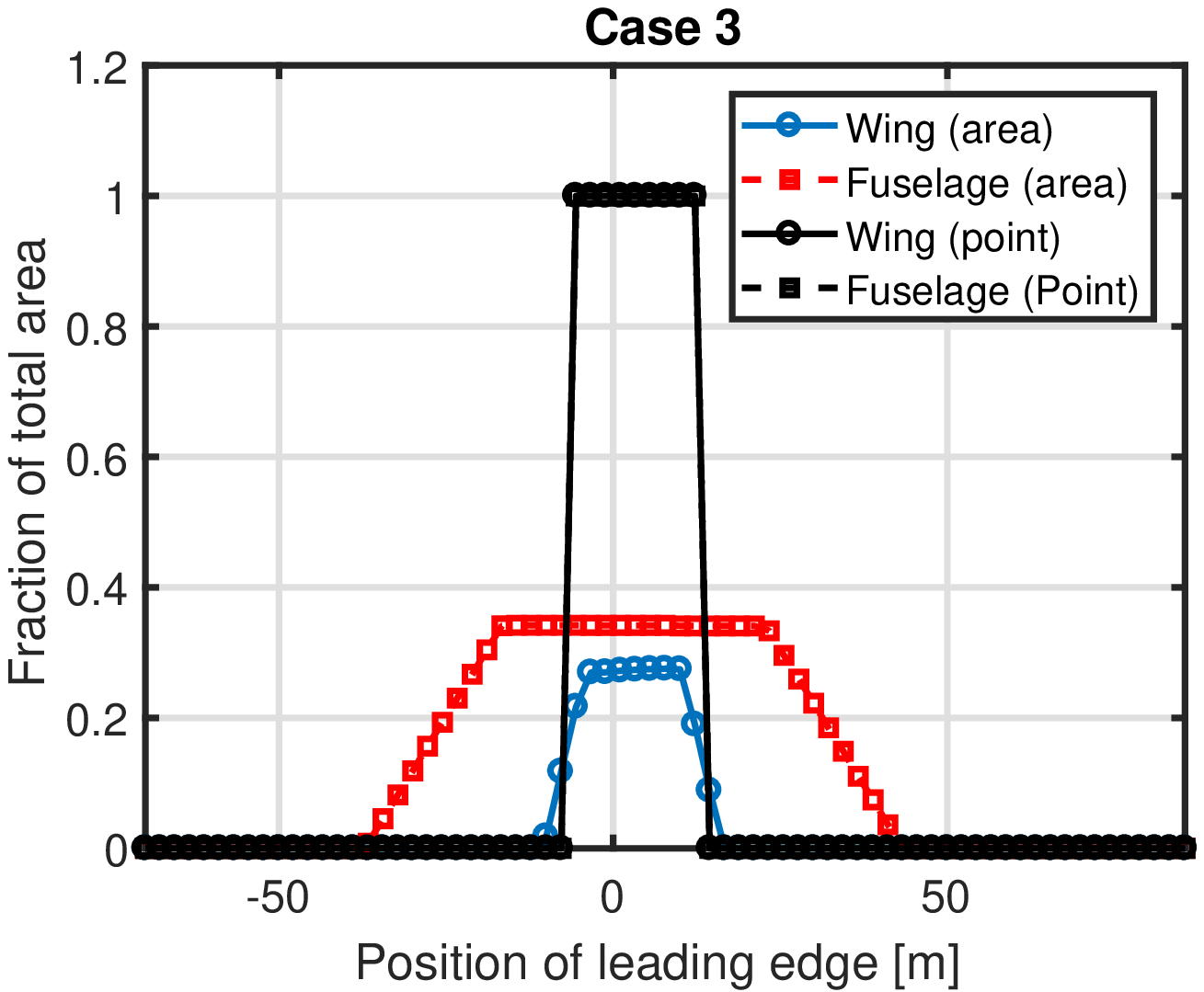}
\caption{Baseline vortex (case 3): Area fraction as function of time (left) and LE position (right). When considering the aircraft as a point, the wing and fuselage area fractions are identical.}
\label{fig:case_3_frac}
\end{figure}

The area fractions are shown both as a function of time and leading edge position for our baseline case in Fig. \ref{fig:case_3_frac}. When considering the aircraft as a point, the area fraction is either zero (aircraft center point outside VT) or one (aircraft center point inside VT). This behaviour is different when the aircraft area is considered. Compared to the wing, the fuselage enters (exits) the VT first (last), respectively. It is also important to note that the area ratios are only about one-third of the point ratios, since the VT only intersects part of the aircraft.

All the cases we have modelled can be found in Table \ref{tab:cases}.

\begin{table}
\caption{Table of cases. Case 3 is the baseline case.}
\label{tab:cases}
 \begin{tabular}{ccccccc}
  \hline
  Case no & $\phi$ [rad] & $\theta$ [rad & $y_0$ [m] & $z_0$ [m] & $R$ [m] & $W$ [m] \\
  \hline
  1  & $\pi/2$ & $\pi/2$ & 0 & 0 & 33.9 & 53.2 \\
  2  & $\pi/2$ & $\pi/2$ & 0 & 0 & 3.4 & 5.3 \\
  3  & $\pi/2$ & $\pi/2$ & 0 & 0 & 10.7 & 16.8 \\
  4  & $\pi/2$ & $\pi/2$ & 8.4 & 0 & 10.7 & 16.8 \\
  5  & $\pi/2$ & $\pi/2$ & 0 & 9.3 & 10.7 & 16.8 \\
  6  & $\pi/4$ & $\pi/2$ & 0 & 0 & 10.7 & 16.8 \\
  7  & $3\pi/4$ & $\pi/2$ & 0 & 0 & 10.7 & 16.8 \\
  8  & $\pi/2$ & $\pi/4$ & 0 & 0 & 10.7 & 16.8 \\
  9  & $\pi/2$ & $3\pi/4$ & 0 & 0 & 10.7 & 16.8 \\
  10 & $\pi/4$ & $\pi/4$ & 0 & 0 & 10.7 & 16.8 \\
  11 & $\pi/4$ & $3\pi/4$ & 0 & 0 & 10.7 & 16.8 \\
  12 & $3\pi/4$ & $\pi/4$ & 0 & 0 & 10.7 & 16.8 \\
  13 & $3\pi/4$ & $3\pi/4$ & 0 & 0 & 10.7 & 16.8 \\
  14 & $0$ & $\pi/2$ & 0 & 0 & 10.7 & 16.8 \\
  15 & $\pi$ & $\pi/2$ & 0 & 0 & 10.7 & 16.8 \\
  16 & $\pi/2$ & $0$ & 0 & 0 & 10.7 & 16.8 \\
  17 & $\pi/2$ & $\pi$ & 0 & 0 & 10.7 & 16.8 \\
  \hline
\end{tabular}
\end{table}

\subsection{Baseline}

\begin{figure}
\includegraphics[width=5cm]{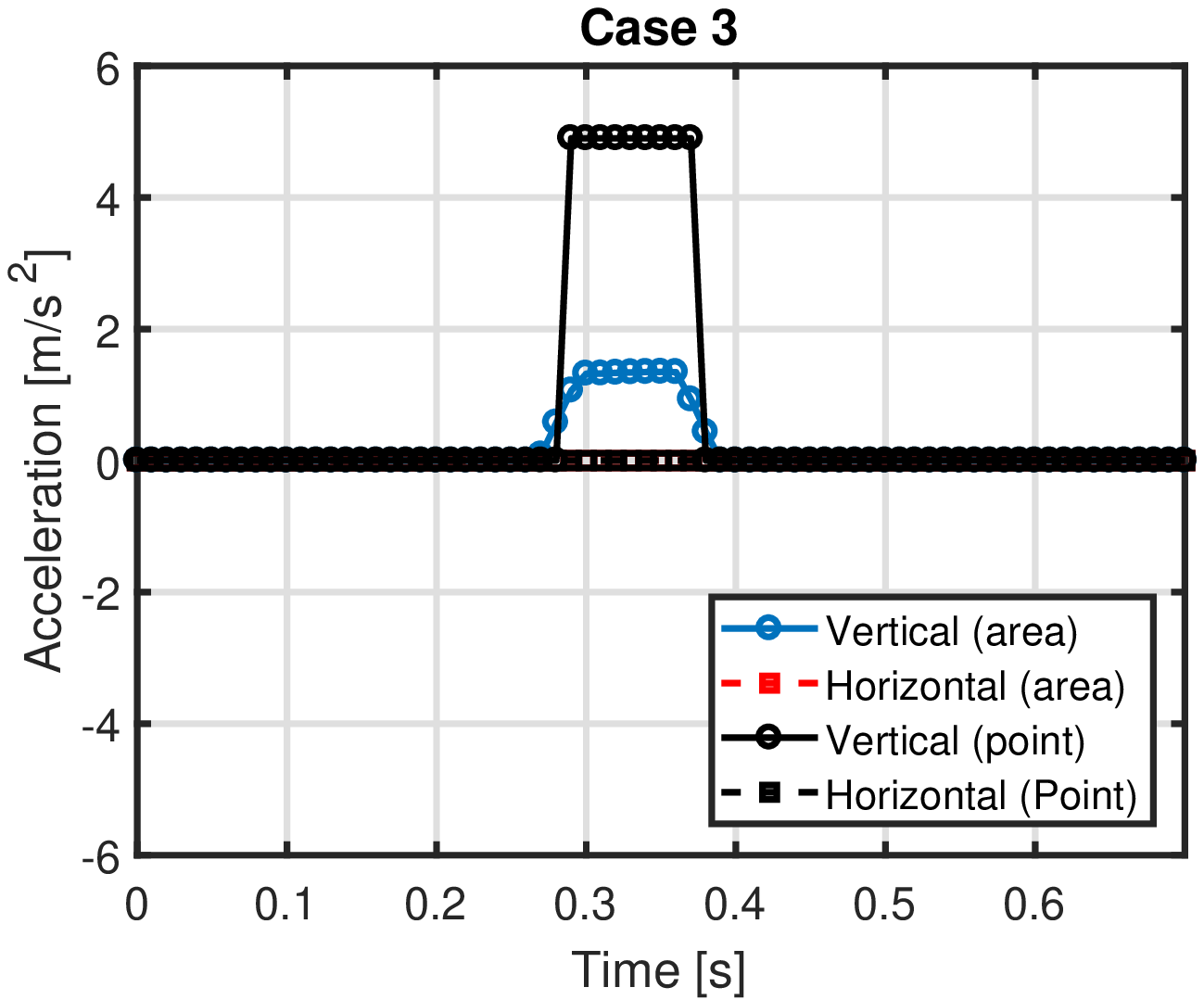}
\hspace{0.5cm}
\includegraphics[width=5cm]{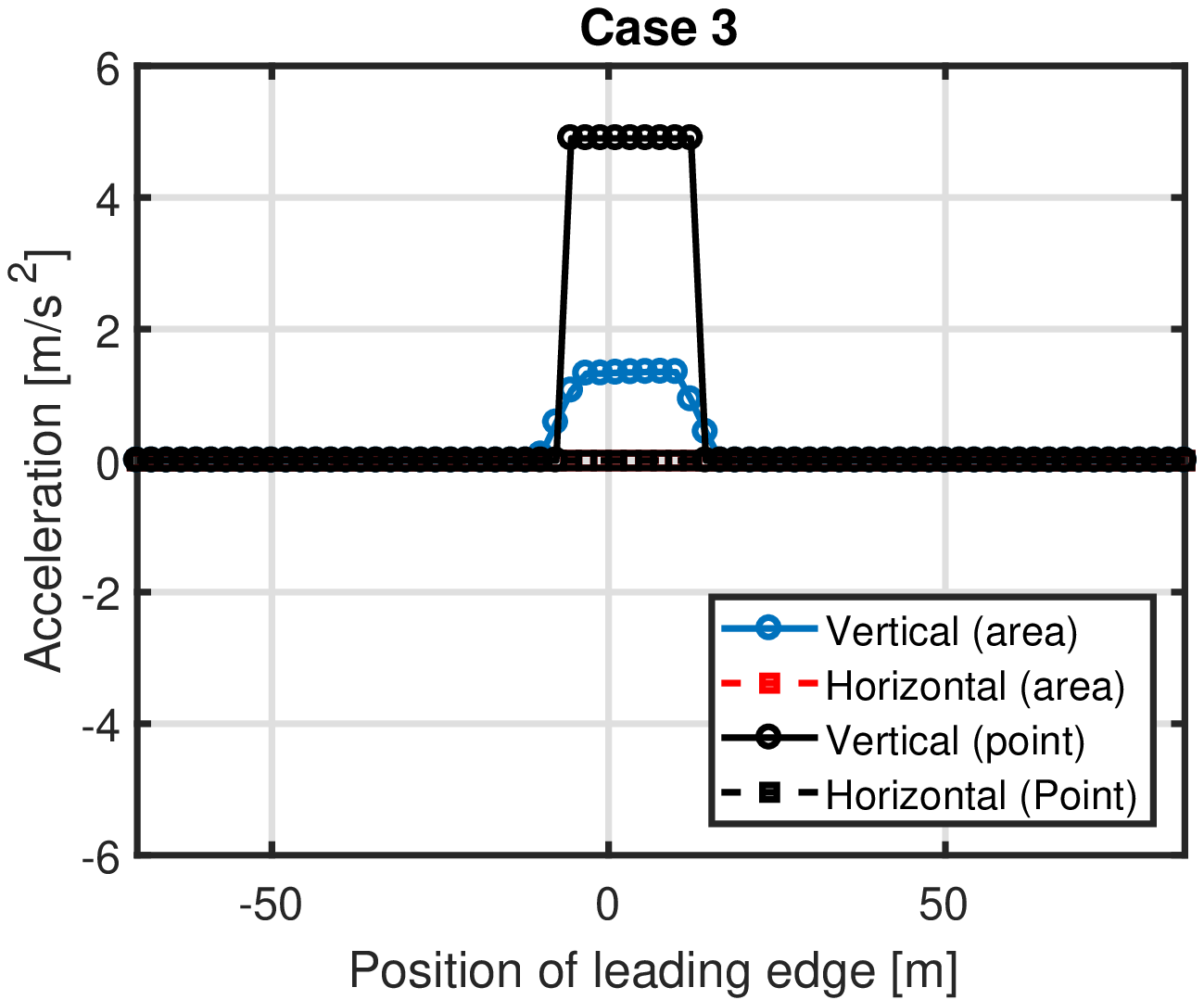}
\caption{Baseline vortex (case 3): Acceleration as function of time (left) and LE position (right). The horizontal acceleration is zero.}
\label{fig:case_3_acc}
\end{figure}

Aircraft accelerations corresponding to the area ratios in Fig. \ref{fig:case_3_frac} are shown in Fig. \ref{fig:case_3_acc}. First of all we note that there is no horizontal acceleration, consistent with our results in Section \ref{sec:ac_as_pt}. The vertical area acceleration is about one-third of the acceleration if one considers the aircraft as a point, see our discussion above. Further, the increase/decrease of acceleration for the vertical acceleration as an area is not instantaneous, since the wing has a finite width, the chord $C$.

This result indicates that it is important to take the aircraft area into account.

For the following analysis, we only show acceleration as a function of time.

\subsection{Size}

Once we have established the baseline case, we can use the modelling procedure to gain insight into the impact of other VT geometries.

\begin{figure}
\includegraphics[width=3.5cm]{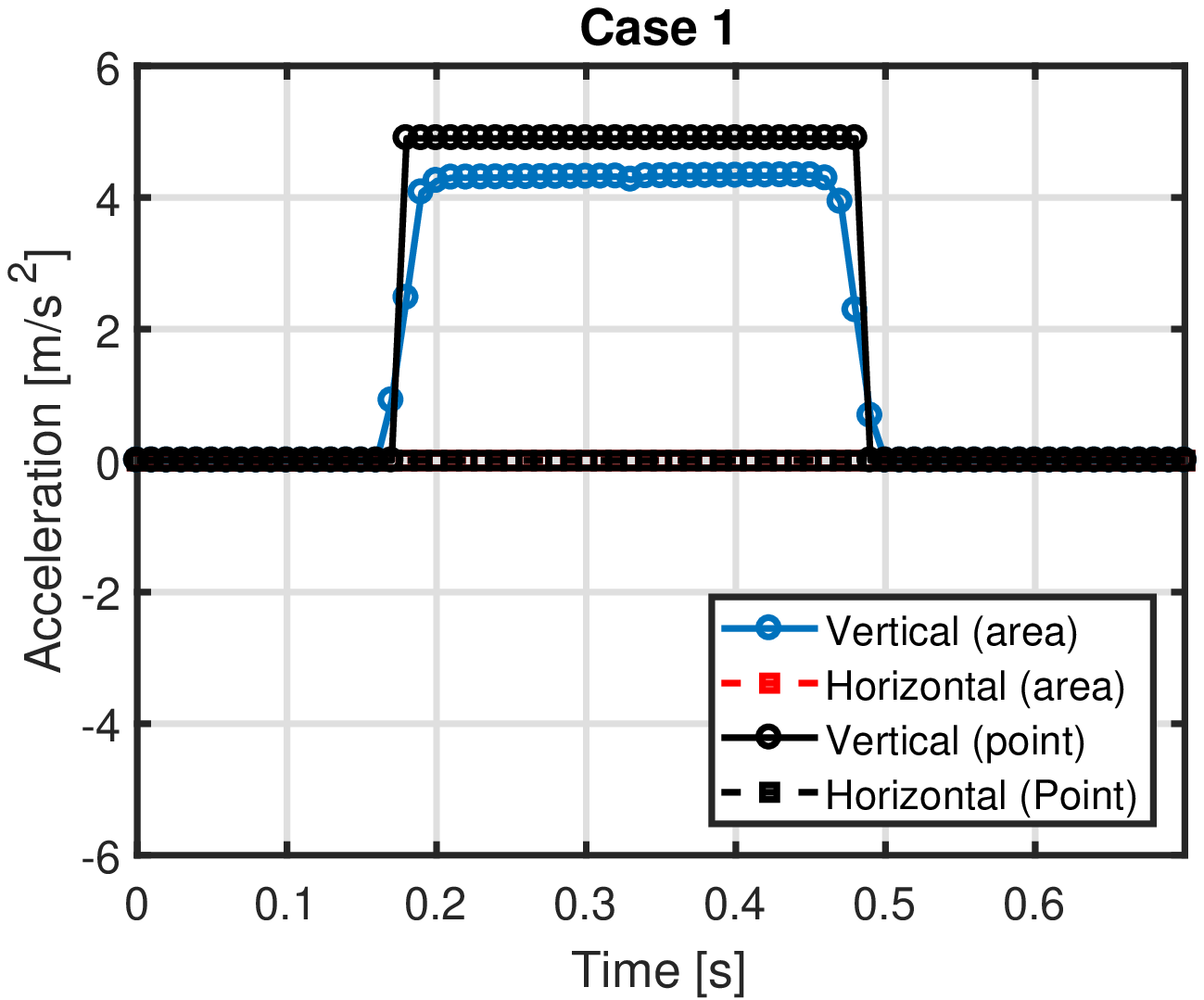}
\hspace{0.5cm}
\includegraphics[width=3.5cm]{case_3_acceleration_time.eps}
\hspace{0.5cm}
\includegraphics[width=3.5cm]{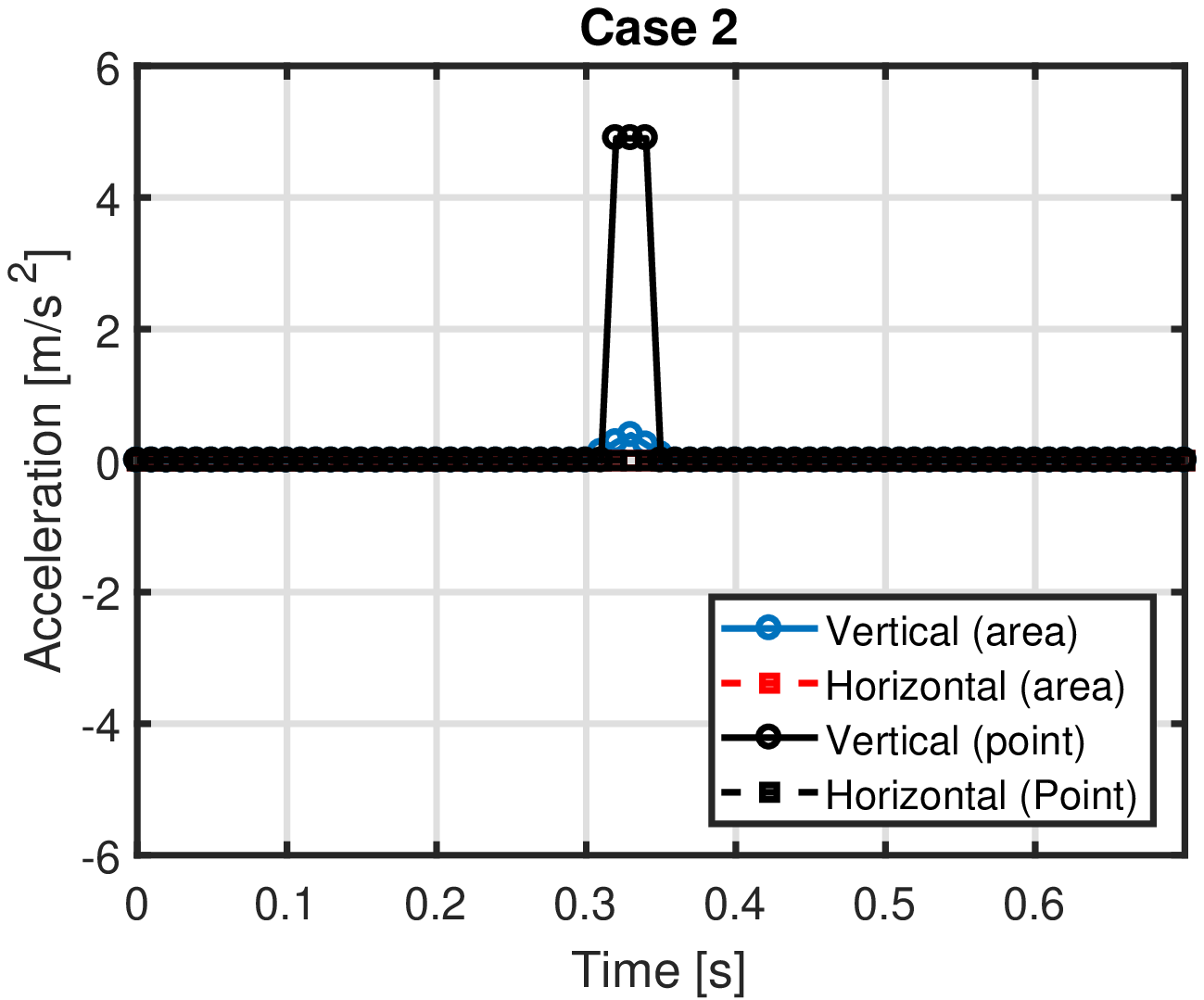}
\caption{Acceleration as function of time for vortex size variation: Large vortex (left), baseline vortex (center), small vortex (right). The horizontal acceleration is zero.}
\label{fig:size_acc}
\end{figure}

First we vary the VT size, see Fig. \ref{fig:size_acc}. These are cases 1, 2 and 3 in Table \ref{tab:cases}. The sizes are determined in Section \ref{subsubsec:p_vs_a}.

As the vortex size decreases, we note that the amplitude of the area acceleration decreases. This confirms that considering the aircraft as a point works well for a large VT, but not for a smaller VT.

\subsection{Offset}

\begin{figure}
\includegraphics[width=3.5cm]{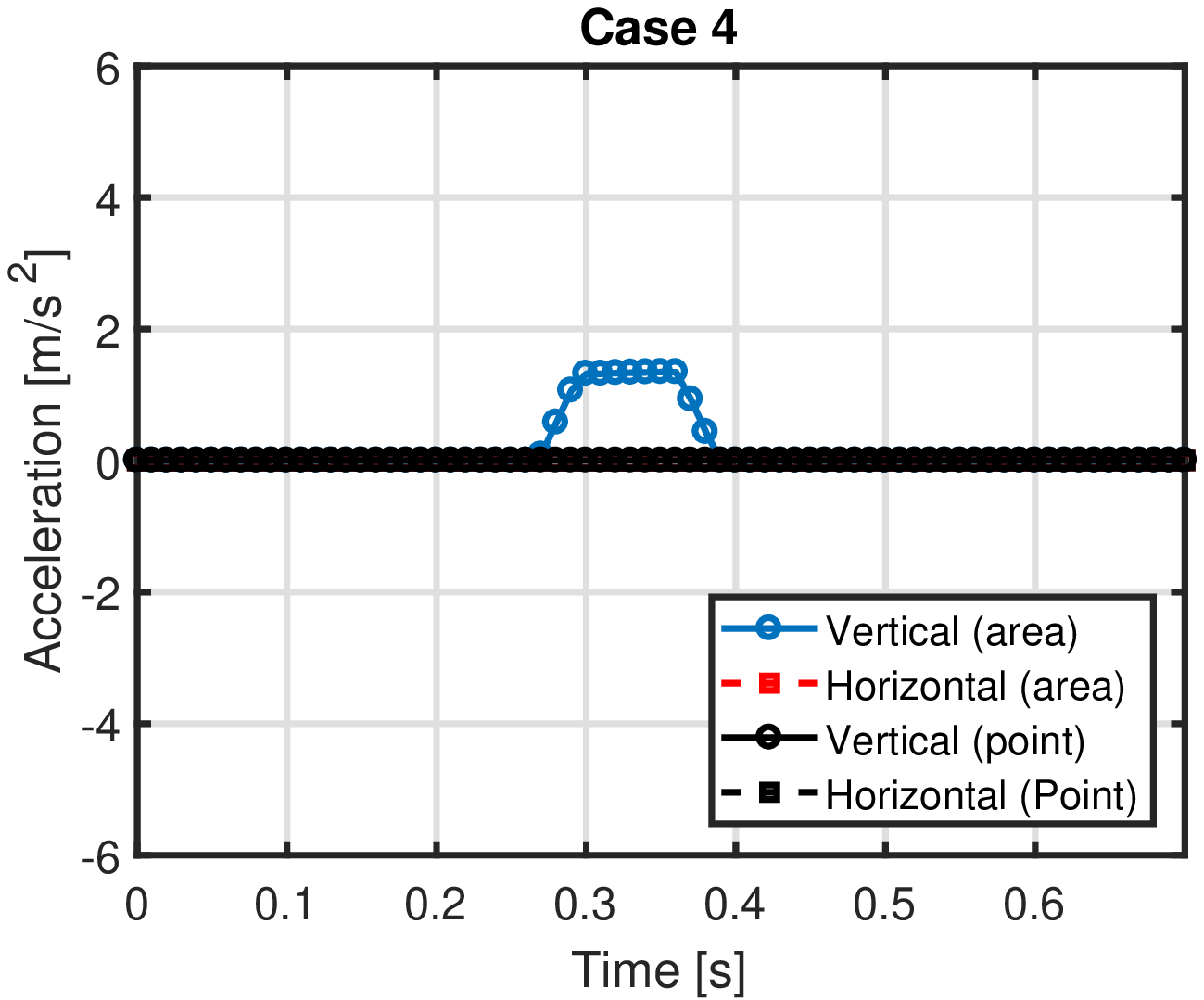}
\hspace{0.5cm}
\includegraphics[width=3.5cm]{case_3_acceleration_time.eps}
\hspace{0.5cm}
\includegraphics[width=3.5cm]{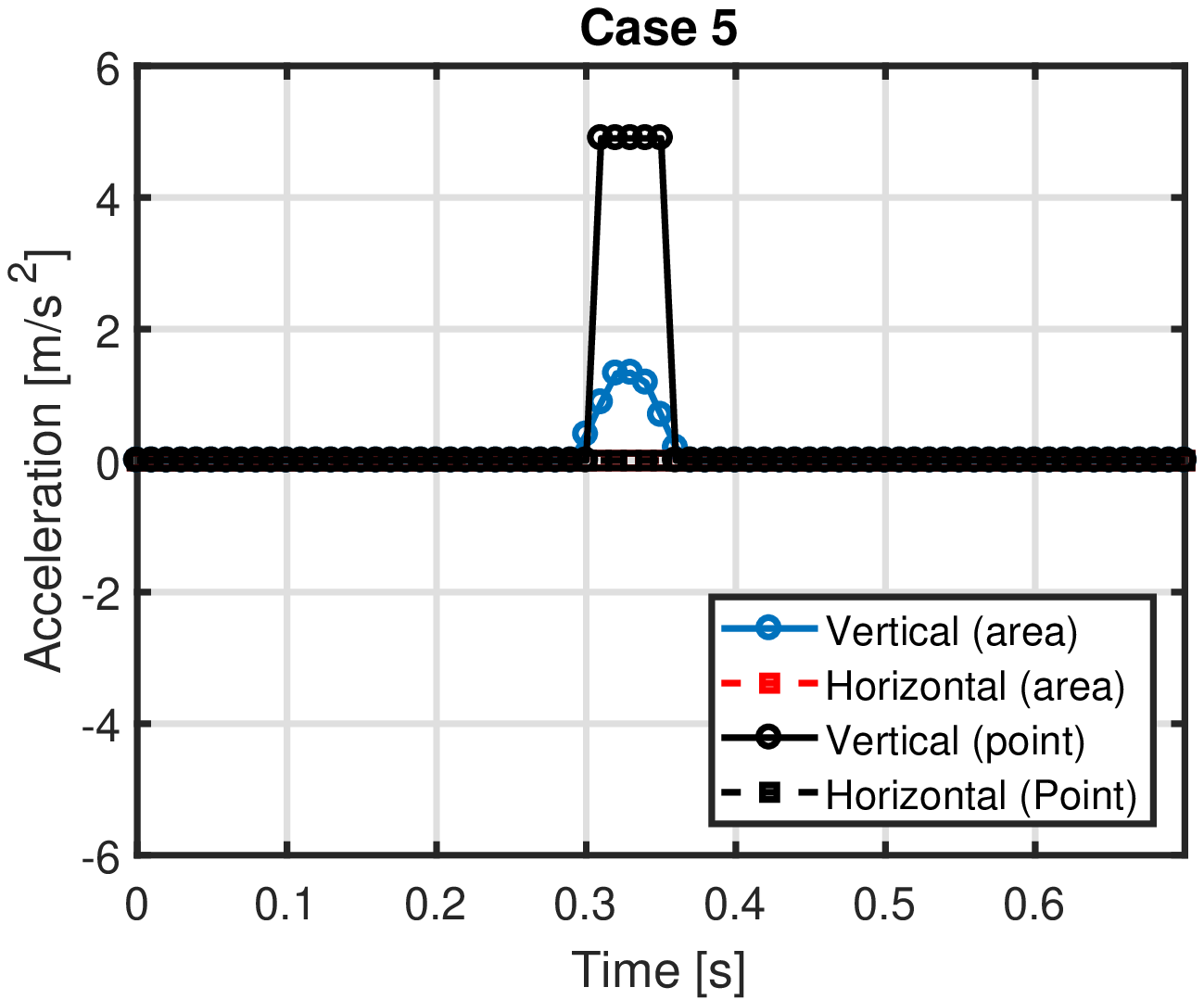}
\caption{Acceleration as function of time for vortex offset variation: Horizontal offset (left), no offset (center), vertical offset (right). The horizontal acceleration is zero.}
\label{fig:offset_acc}
\end{figure}

We continue the modelling by investigating the effect of an offset VT, see Fig. \ref{fig:offset_acc}. These are cases 3, 4 and 5. Moving the VT by $R$ in the horizontal direction means that the point acceleration is zero, i.e. the aircraft center point is exactly at the edge of the VT. A vertical VT offset by $R\sqrt{\frac{3}{4}}$ (see Section \ref{subsec:vt_offset}) means that the VT extent in the $x$-direction is reduced from $2R$ to $R$. This decrease manifests itself as a shorter duration of the aircraft-VT encounter.

\subsection{Angle: $\theta=\pi/2$, $\phi$ Varies}

\begin{figure}
\includegraphics[width=3.5cm]{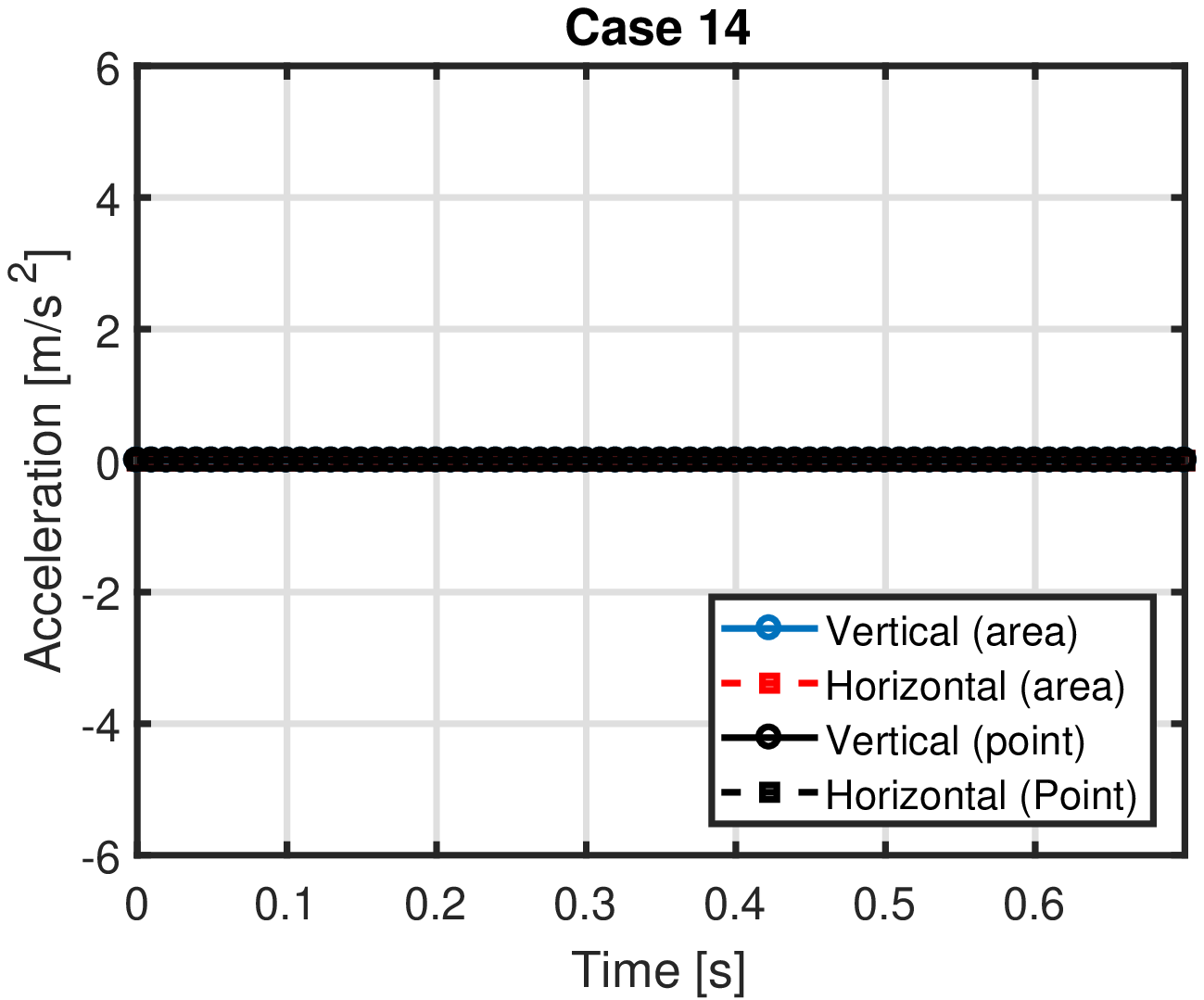}
\hspace{0.5cm}
\includegraphics[width=3.5cm]{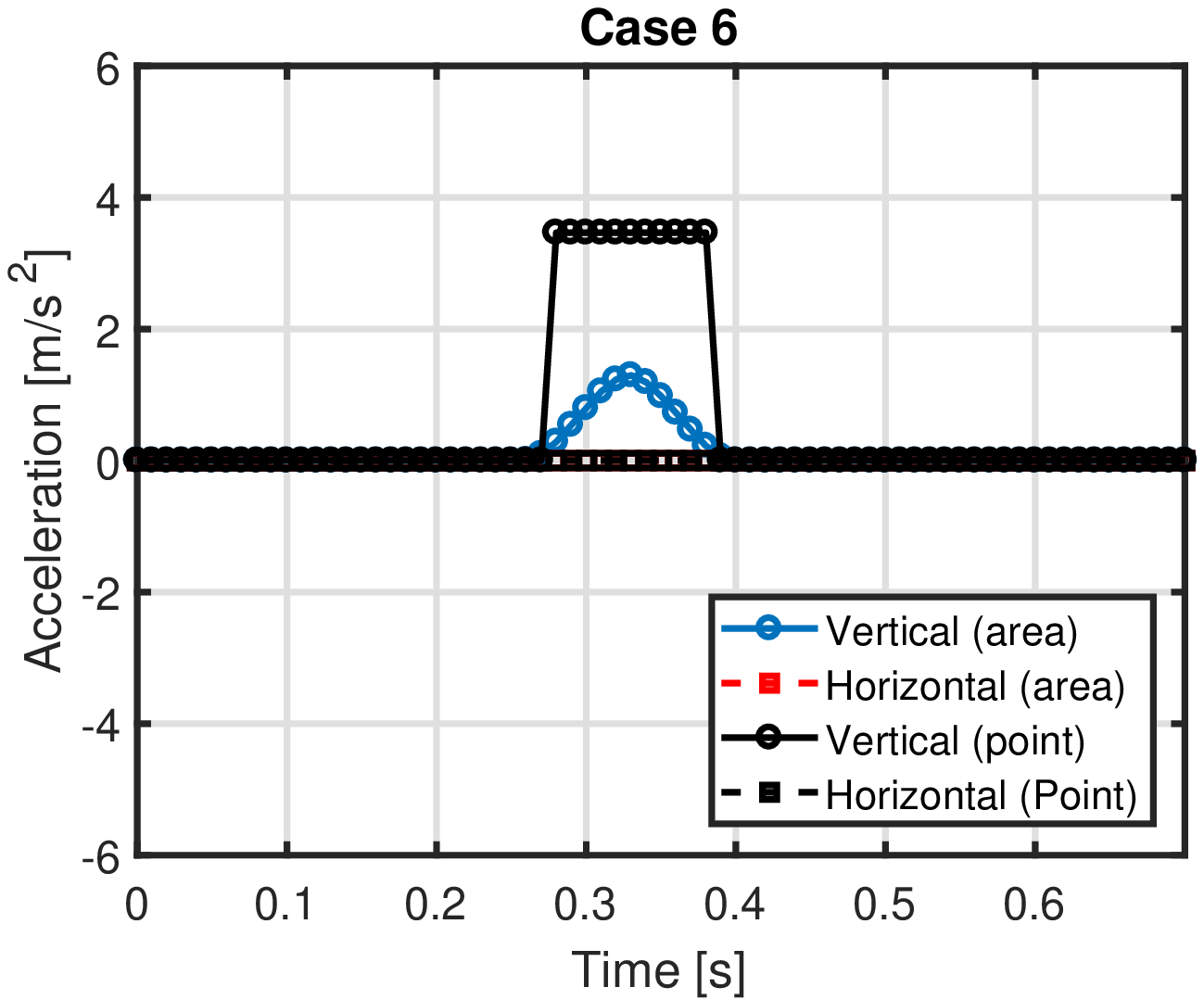}
\hspace{0.5cm}
\includegraphics[width=3.5cm]{case_3_acceleration_time.eps}
\caption{Acceleration as function of time for vortex $\phi$ variation: $\phi=0$ or $\phi=\pi$ (left), $\phi=\pi/4$ or $\phi=3\pi/4$ (center), $\phi=\pi/2$ (right). The horizontal acceleration is zero.}
\label{fig:phi_vary_acc_small}
\end{figure}

Here, we present results for the left-right tilt of the VT using the angle $\phi$, see Fig. \ref{fig:phi_vary_acc_small}. There is no up-down tilt, $\theta$ is kept fixed at $\pi/2$. These are cases 3, 6 and 14.

If the VT is parallel to the aircraft, there is no acceleration. At an angle $\phi=\pi/4$ (or $3\pi/4$), the area acceleration does not have a flat top and the duration of the aircraft-VT encounter is longer.

\subsection{Angle: $\phi=\pi/2$, $\theta$ Varies}

For the up-down tilt, there is no left-right tilt, i.e. $\phi=\pi/2$.

\begin{figure}
\includegraphics[width=3.5cm]{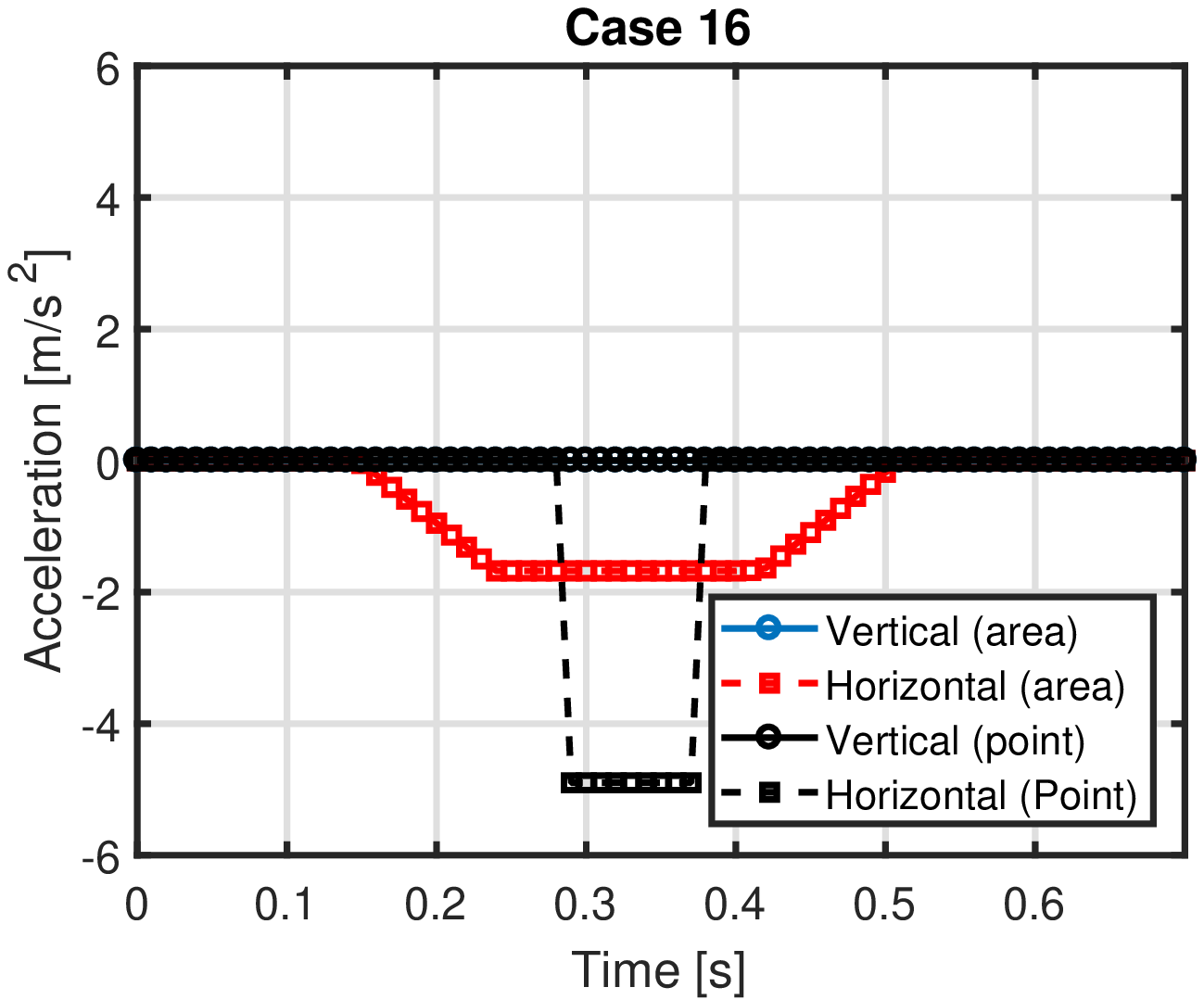}
\hspace{0.5cm}
\includegraphics[width=3.5cm]{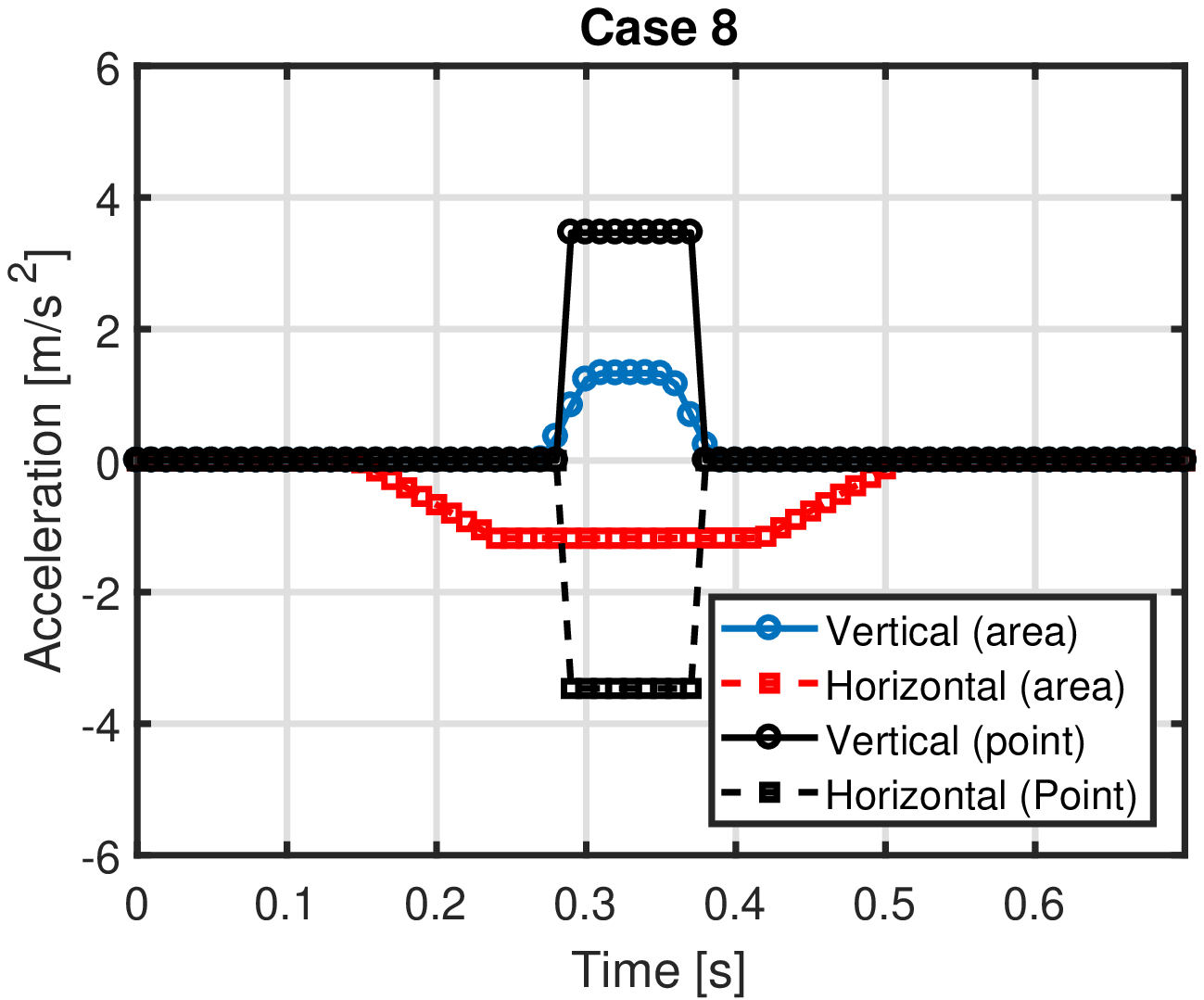}
\hspace{0.5cm}
\includegraphics[width=3.5cm]{case_3_acceleration_time.eps}
\caption{Acceleration as function of time for vortex $\theta$ variation: $\theta=0$ (left, no vertical acceleration), $\theta=\pi/4$ (center), $\theta=\pi/2$ (right, no horizontal acceleration).}
\label{fig:theta_vary_acc_small}
\end{figure}

We first present the results for $\theta$ less than $\pi/2$, see Fig. \ref{fig:theta_vary_acc_small}. These are cases 3, 8 and 16. An angle $\theta=0$ corresponds to a columnar vortex; here, only a horizontal acceleration takes place. The amplitude is lower and duration longer than when modelling the aircraft as a point. Both horizontal and vertical accelerations are present for $\theta=\pi/4$.

\begin{figure}
\includegraphics[width=3.5cm]{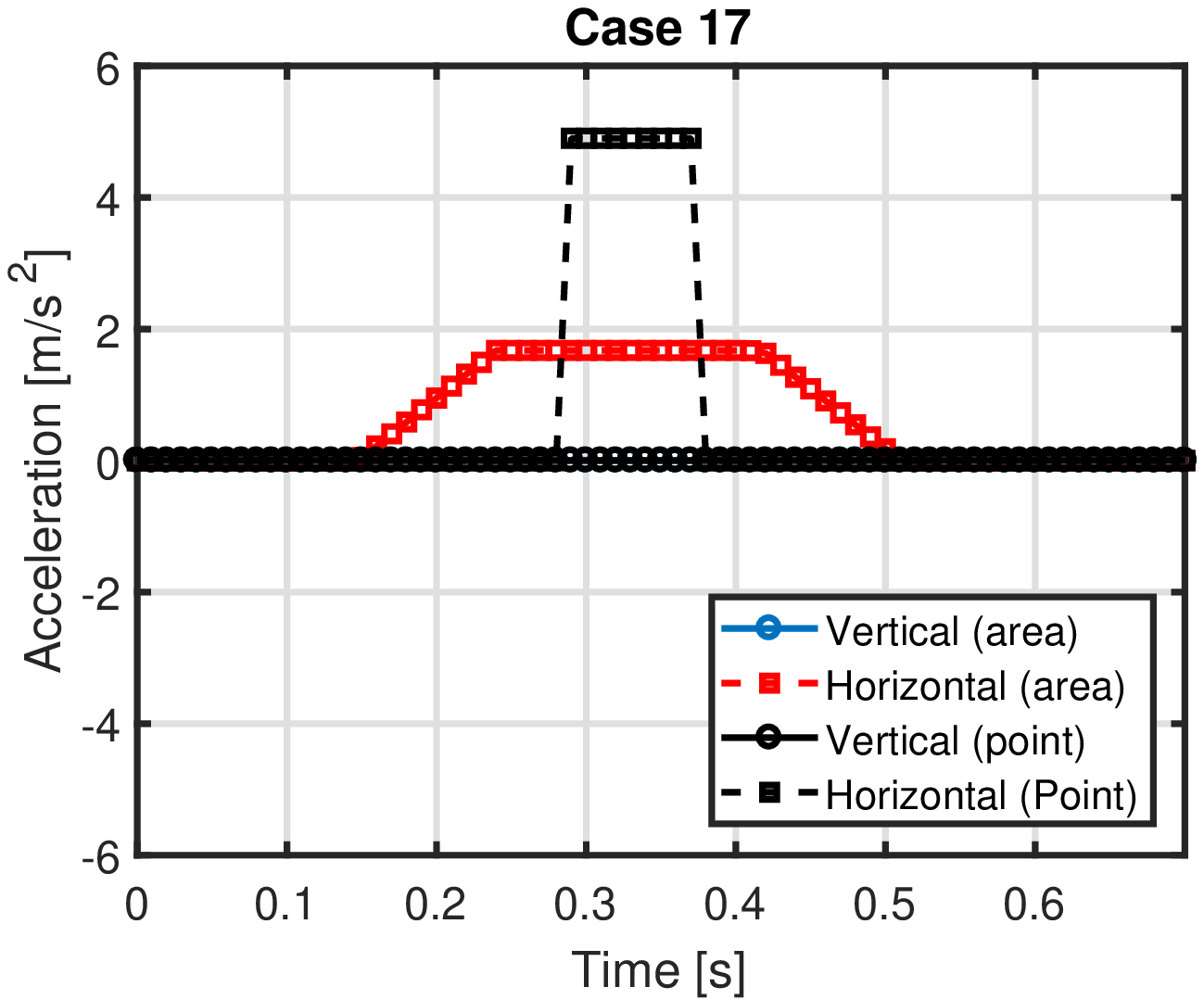}
\hspace{0.5cm}
\includegraphics[width=3.5cm]{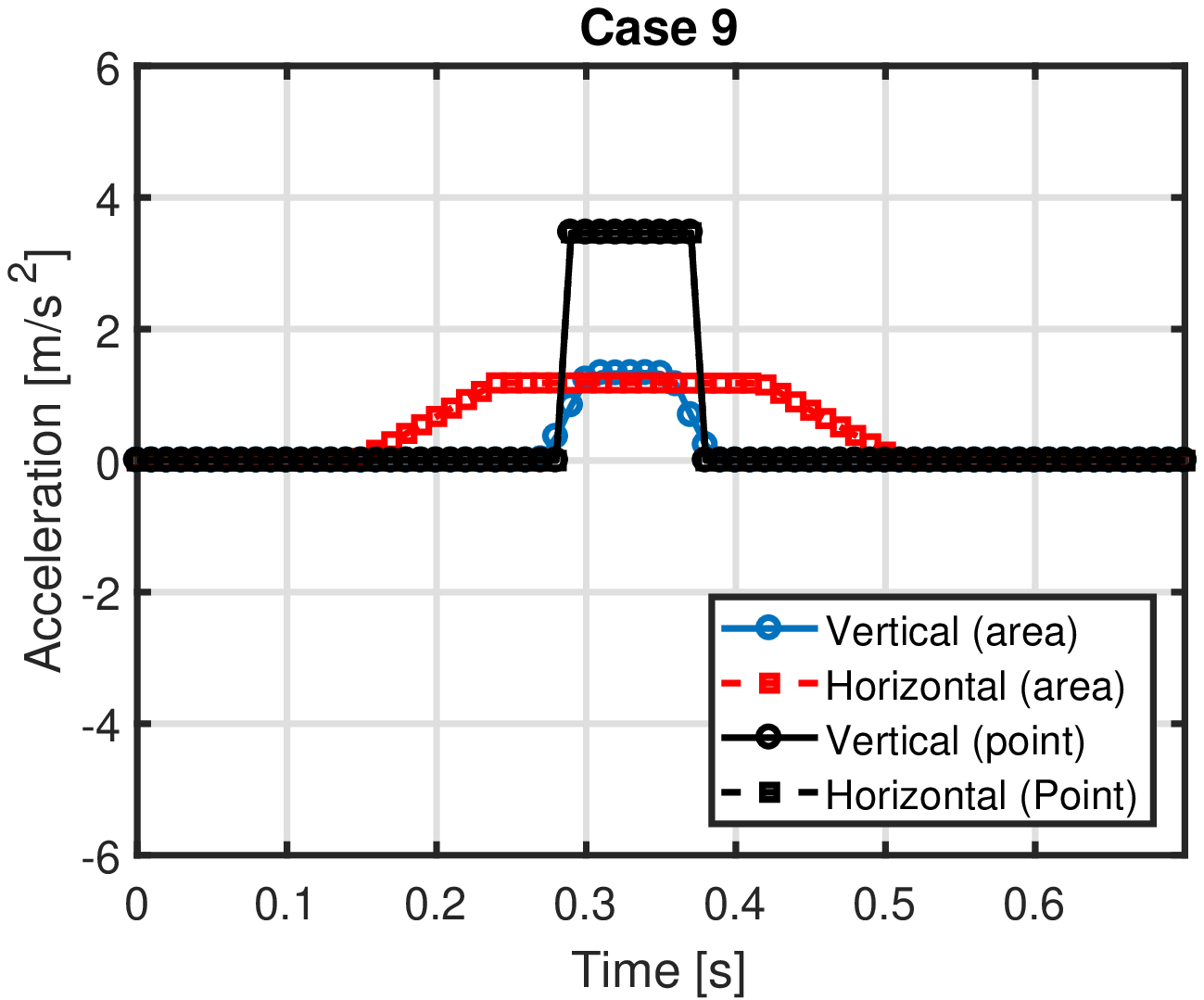}
\hspace{0.5cm}
\includegraphics[width=3.5cm]{case_3_acceleration_time.eps}
\caption{Acceleration as function of time for vortex $\theta$ variation: $\theta=\pi$ (left, no vertical acceleration), $\theta=3\pi/4$ (center), $\theta=\pi/2$ (right, no horizontal acceleration). For $\theta=3\pi/4$ (center), the vertical and horizontal acceleration is identical when considering the aircraft as a point.}
\label{fig:theta_vary_acc_large}
\end{figure}

We now proceed to present the results for $\theta$ larger than $\pi/2$, see Fig. \ref{fig:theta_vary_acc_large}. These are cases 3, 9 and 17. The vertical acceleration has the same amplitude and direction as for the smaller angles. The horizontal acceleration has the same amplitude but the opposite sign.

\subsection{Angle: Both Angles Vary}

\begin{figure}
\includegraphics[width=3.5cm]{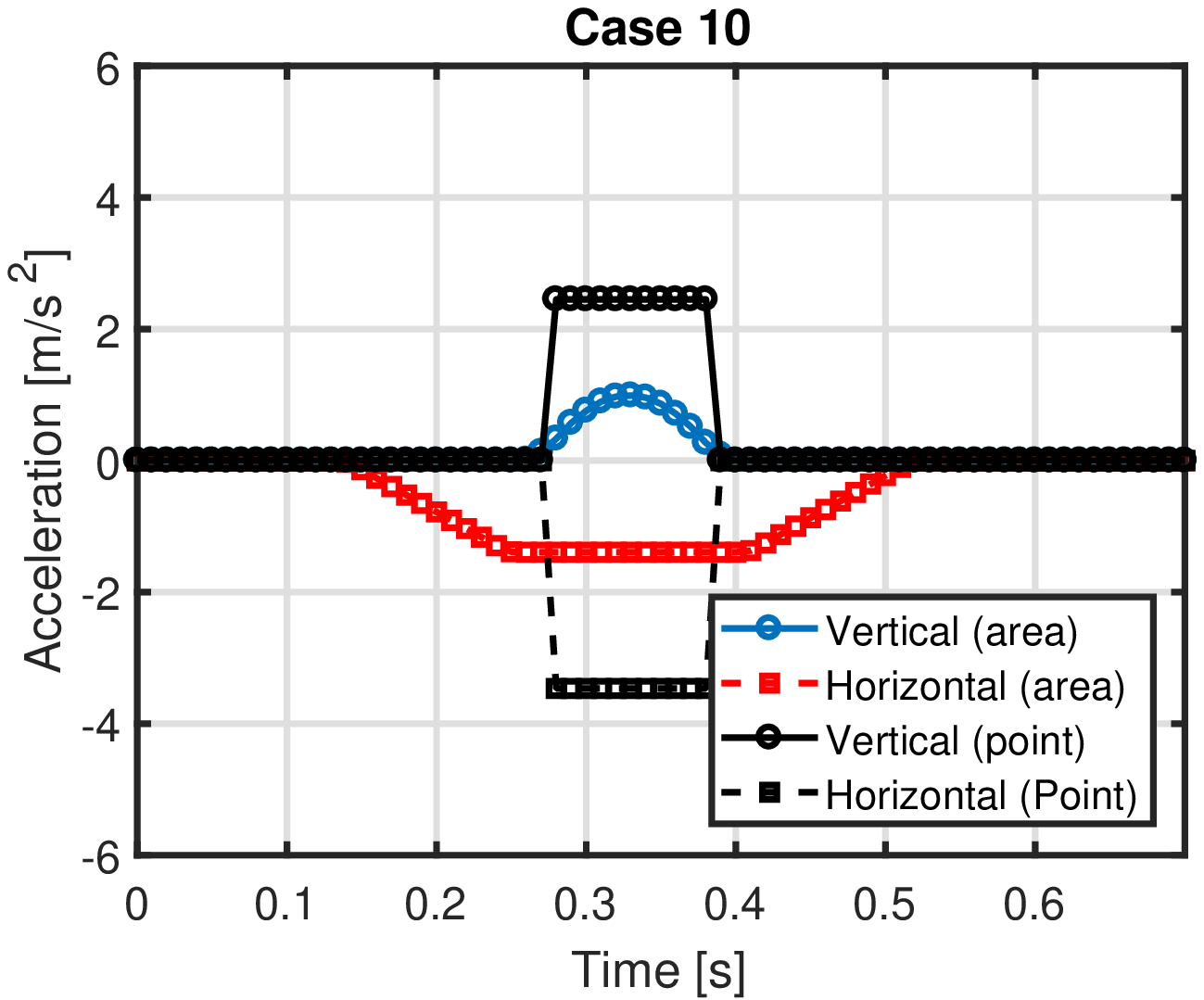}
\hspace{0.5cm}
\includegraphics[width=3.5cm]{case_3_acceleration_time.eps}
\hspace{0.5cm}
\includegraphics[width=3.5cm]{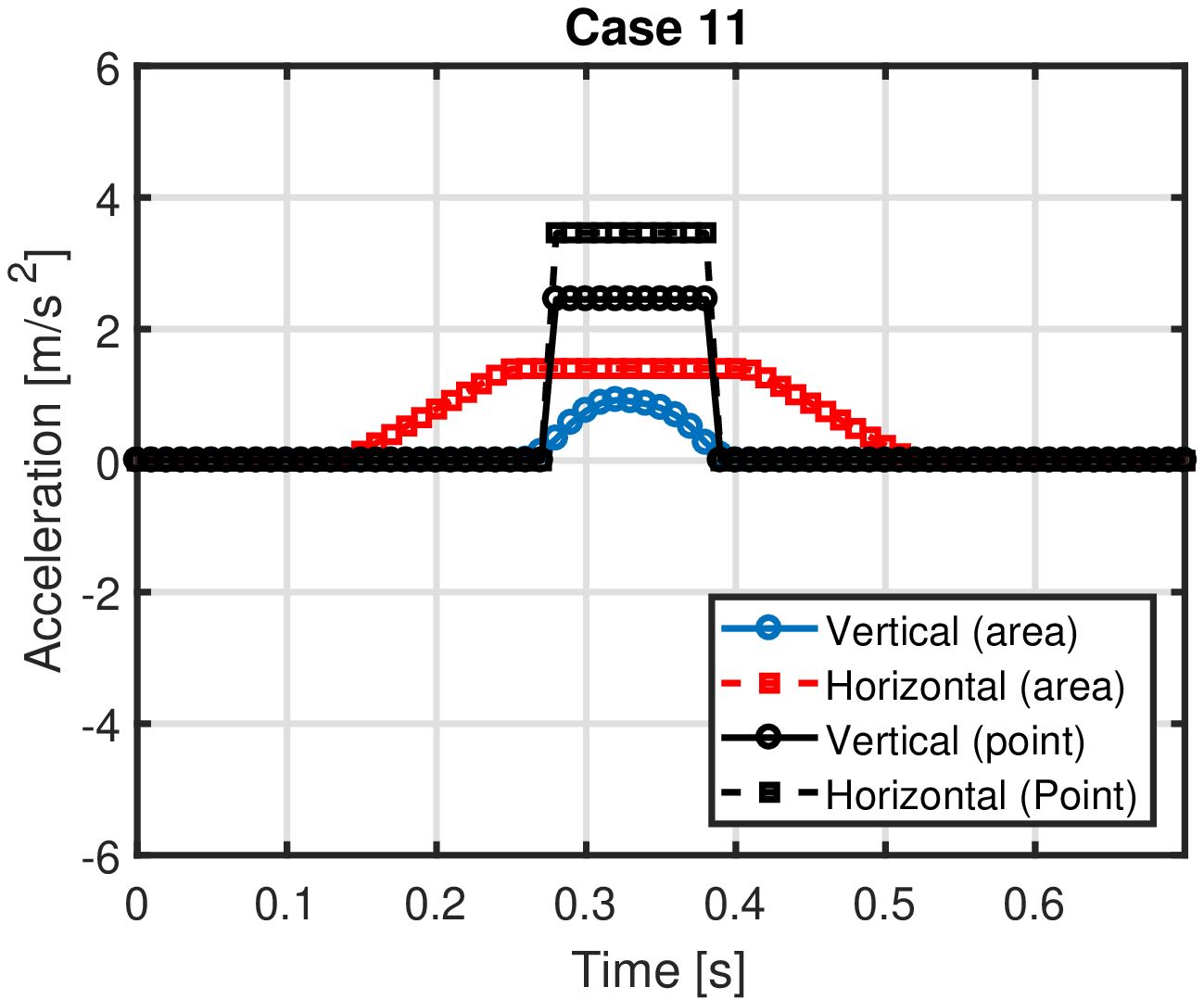}
\caption{Acceleration as function of time for vortex combined angle variation: $\phi=\pi/4, \theta=\pi/4$ or $\phi=3\pi/4, \theta=\pi/4$ (left), $\phi=\pi/2, \theta=\pi/2$ (center, no horizontal acceleration), $\phi=\pi/4, \theta=3\pi/4$ or $\phi=3\pi/4, \theta=3\pi/4$ (right).}
\label{fig:both_vary_acc_lo_phi}
\end{figure}

Removing the restricting of a fixed $\phi$ or $\theta$ we can consider other combinations, see Fig. \ref{fig:both_vary_acc_lo_phi}. For these cases, the amplitude of the vertical and horizontal acceleration differs.

\section{Discussion and Future Work}

The model we have presented above is a starting point; it is relatively straightforward to extend our model.

\begin{figure}
\includegraphics[width=3.5cm]{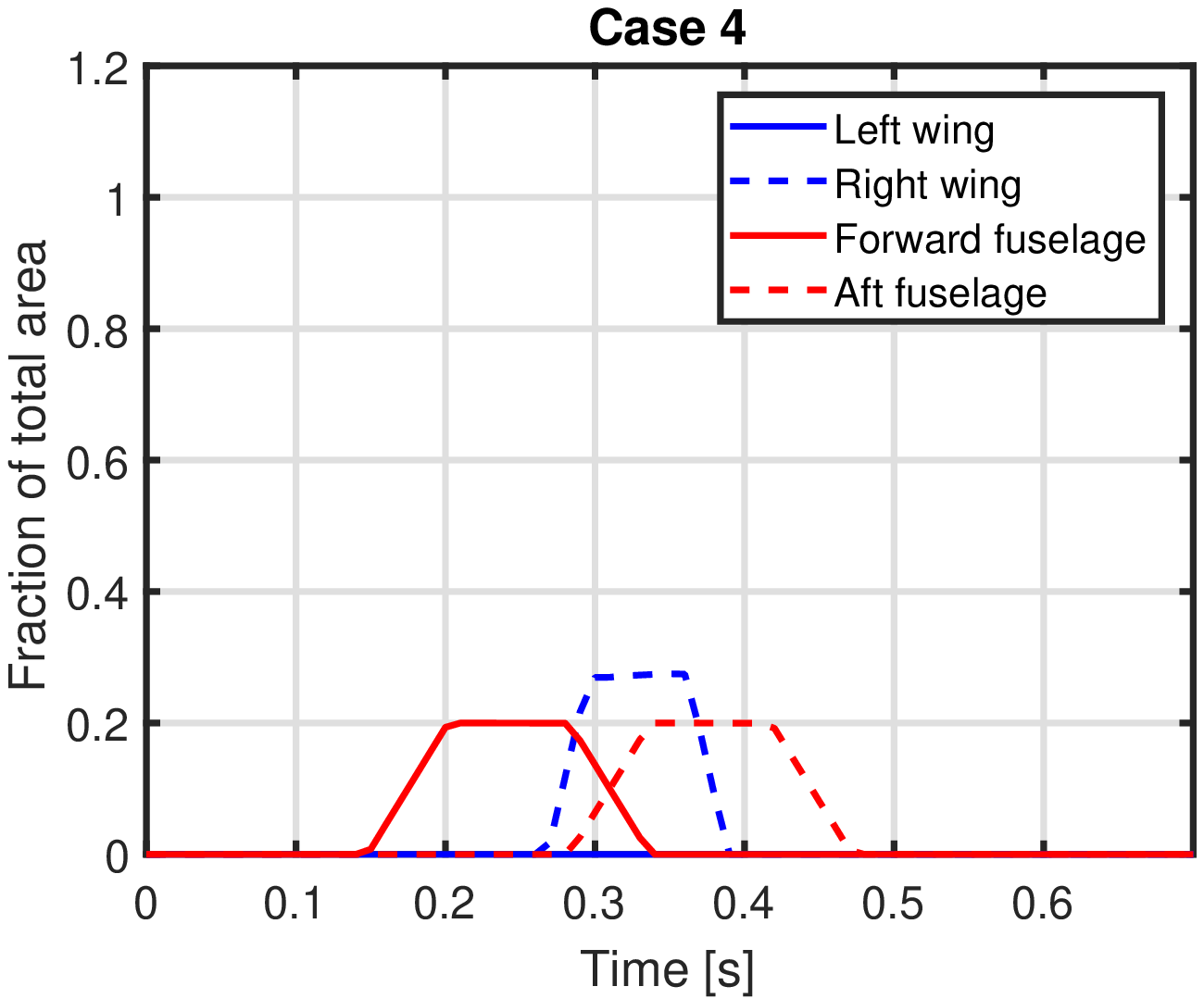}
\hspace{0.5cm}
\includegraphics[width=3.5cm]{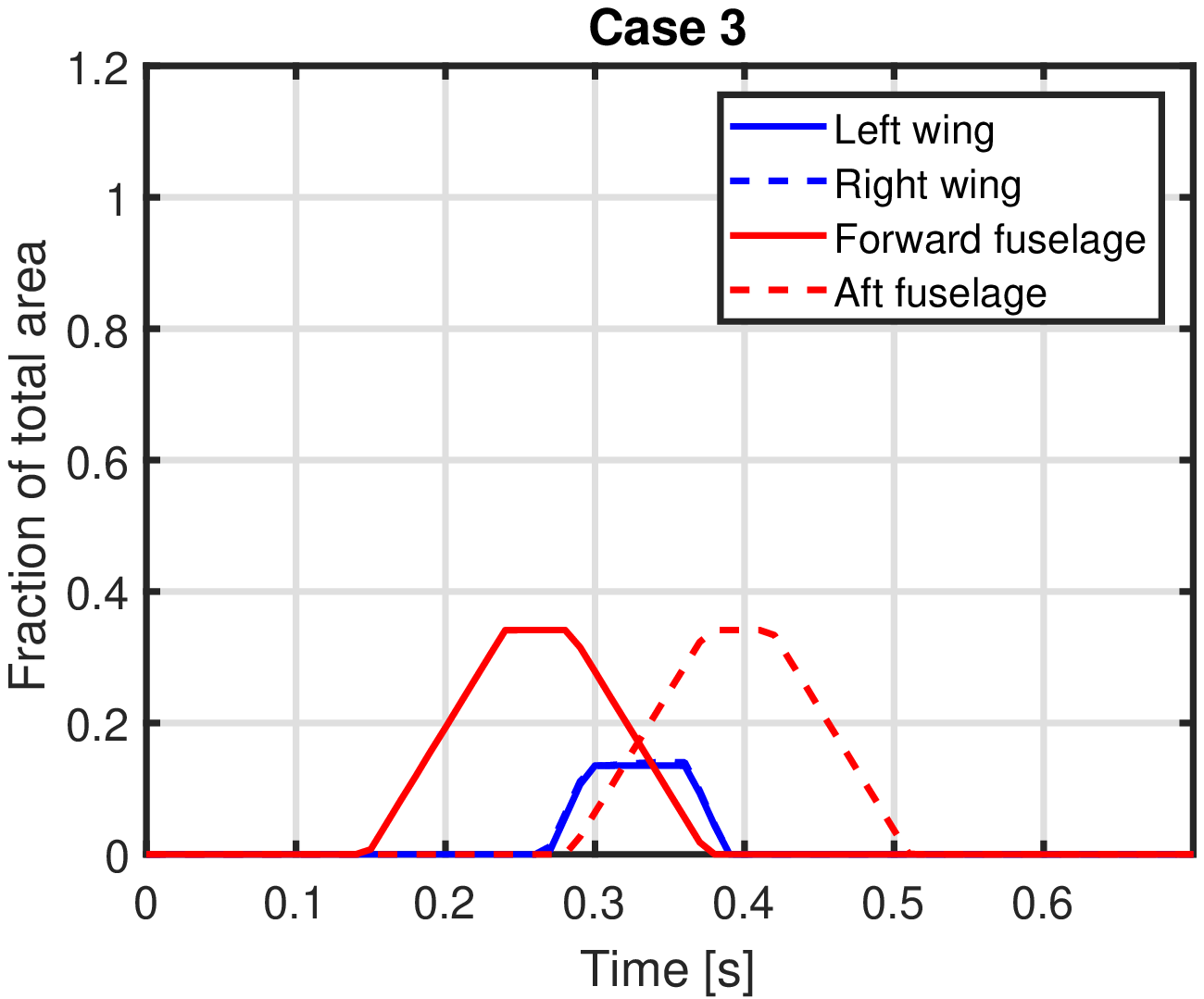}
\hspace{0.5cm}
\includegraphics[width=3.5cm]{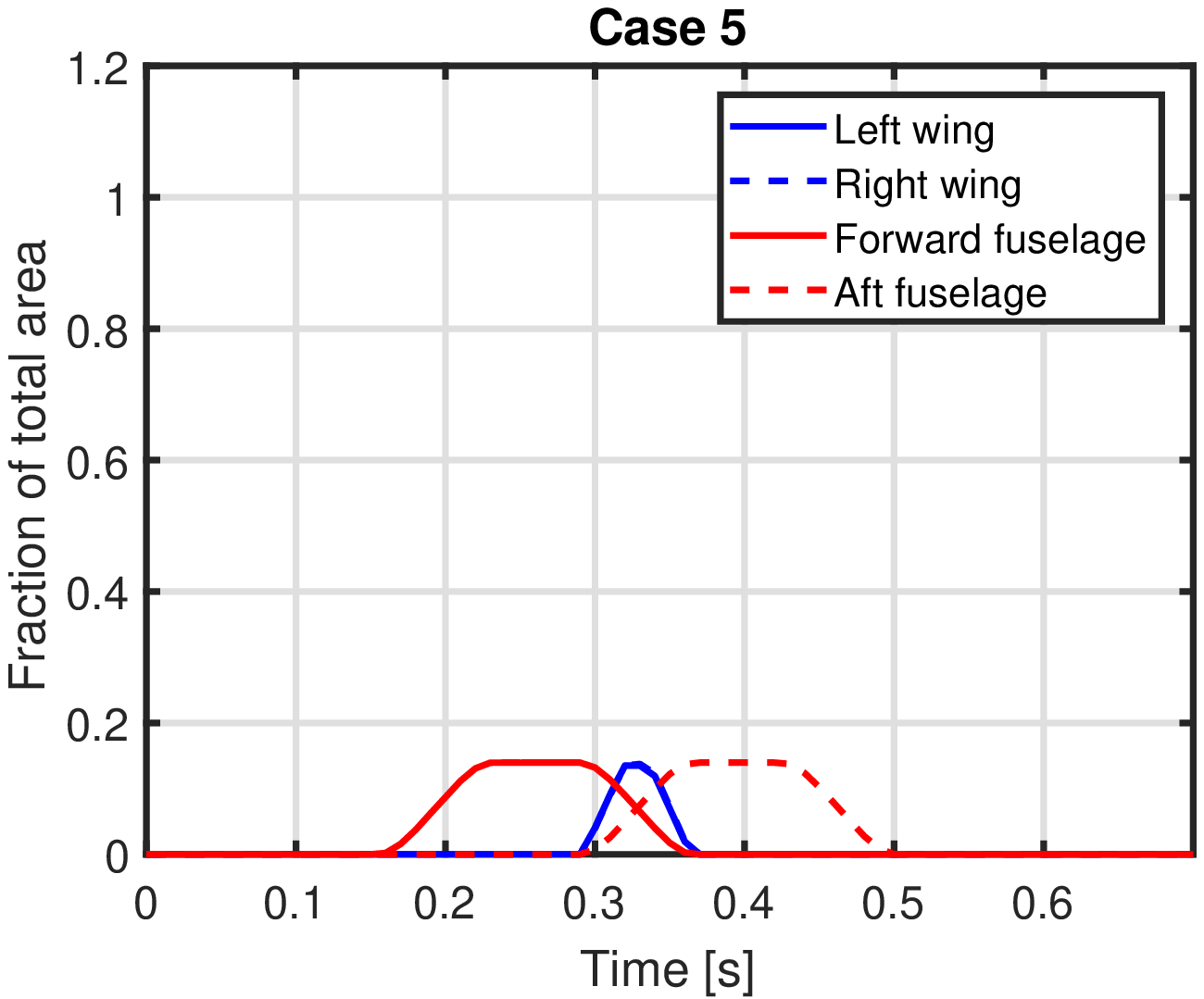}
\caption{Area fraction as function of time:  Horizontal offset (left), no offset (center), vertical offset (right). For the left-hand plot, the area fraction of the left wing is zero; for the center and right-hand plots, the left and right wing ratios are identical.}
\label{fig:offset_wing_fuselage}
\end{figure}

One example is asymmetry effects: The left/right wing and forward/aft fuselage can be treated separately. As an illustration we show the offset cases (4 and 5) in Fig. \ref{fig:offset_wing_fuselage}. Because of the fuselage length, it seems likely that there will be significant stress between the forward and aft part compared to the left and right wing.

The procedure to consider the aircraft divided into smaller sections is also employed in the literature on wake vortices as the strip model \citep{hahn_a}. This paper also includes measurements of vertical and horizontal acceleration when an aircraft passes through the counter-rotating wake vortices generated by another aircraft.

As discussed below, there is a hierarchy of models of both aircraft and vortices and in this paper the models presented are at the simple end. The advantage of such an approach is that the results are relatively straightforward to interpret. Future work should include comparisons of relatively simple models such as those described in the current paper with more complex models (of both aircraft and vortices) so that the effect of the extra complexity can be both quantified and understood.

Several vortex models have been employed for wake vortices \citep{gerz_a,ahmad_a}. They typically consist of one (or two) cores which contain vorticity surrounded by an irrotational outer flow region. Such vortex models could also be studied using our methodology. In \citep{spilman_a}, the Lamb-Oseen vortex model was used for an aircraft on final approach to landing.

Multiple co-rotating vortices have been modelled \citep{parks_a,mehta_a}, and counter-rotating vortices have been identified as well \citep{misaka_a}.

It would be interesting for us to obtain high-quality aircraft acceleration measurements to make a fit to our model with several co- and/or counter-rotating vortices.

A more complete description of the aircraft motion would include e.g. lift and drag forces as presented in \citep{fischenberg_a,schwarz_a}. Depending on the response of the aircraft, it may spend more or less time inside the vortex tube than our model indicates.

\section{Conclusions}

We have extended the modelling of an interaction between an aircraft and a vortex tube to include:

\begin{itemize}
\item An arbitrary vortex tube angle
\item Both vertical and horizontal acceleration
\item An aircraft as having an area (wing and fuselage)
\end{itemize}

Our model shows that modelling the aircraft as a point will not give a correct acceleration amplitude or duration for cases where the vortex area is of the order of the aircraft area or smaller.

We find it plausible that our model can be used to describe both vortices creating clear-air turbulence (CAT) and wake vortices generated by other aircraft.

\section*{Acknowledgements}

We thank a reviewer for providing constructive comments and suggestions, leading to several improvements of the paper.

\section*{Conflict of interest}

The author declares that he has no conflict of interest.


\bibliographystyle{spbasic}      
\bibliography{Basse_MAAP_v4}

\end{document}